\documentclass[12pt]{article}
\pdfoutput=1

\usepackage{cancel,slashed}
\usepackage{bbm}
\usepackage{bm}
\usepackage{amsmath,amsfonts,amssymb,mathtools,nicefrac,nccmath,cases}
\usepackage{graphicx}
\usepackage{cite}
\usepackage{dsfont}
\usepackage{latexsym}
\usepackage{color}
\usepackage[hyperfootnotes=false,linktocpage]{hyperref}
\usepackage{transparent}
\usepackage[hang,flushmargin]{footmisc}
\usepackage{blkarray}
\usepackage{multirow}
\usepackage{empheq}
\usepackage{comment}
\usepackage{wasysym}
\usepackage{tikzsymbols}
\usepackage{caption}
\usepackage{subcaption}
\usepackage[normalem]{ulem}
\usepackage{braket}
\usepackage{float}
\usepackage{enumitem}

\DeclareFontFamily{U}{rcjhbltx}{}
\DeclareFontShape{U}{rcjhbltx}{m}{n}{<->rcjhbltx}{}
\DeclareSymbolFont{hebrewletters}{U}{rcjhbltx}{m}{n}

\begin{document}
	\allowdisplaybreaks
	\pagestyle{plain}
	
	\makeatletter
	\@addtoreset{equation}{section}
	\makeatother
	\renewcommand{\theequation}{\thesection.\arabic{equation}}

	\pagestyle{empty}
	\rightline{IFT-UAM/CSIC-23-108}
	\vspace{0.5cm}
	\begin{center}
		\Huge{{$\alpha'$ corrections to 4-dimensional \\   non-extremal stringy black holes} 
			\\[12mm]}
		\normalsize{Matteo Zatti}\,$^{1,2}$\\[12mm]
		\small{
			${}^{1}$ Instituto de F\'{\i}sica Te\'orica UAM-CSIC, c/ Nicol\'as Cabrera 13-15, 28049 Madrid, Spain \\[2mm] 
			${}^{2}$ Departamento de F\'{\i}sica Te\'orica, Universidad Aut\'onoma de Madrid, 28049 Madrid, Spain
			\\[10mm]} 
		\small{\bf Abstract} \\[5mm]
	\end{center}
	\begin{center}
		\begin{minipage}[h]{15.0cm} 	
		We compute the first-order $\alpha'$ corrections to a family of 4-dimensional, 4-charge, non-extremal black hole solutions of Heterotic Supergravity in the case with 3 independent charges. The solutions are fully analytic, reproduce the extremal limit previously found in the literature and, applying T-duality, they transform as expected. If we reduce to the case with a single independent charge we obtain the corrections to four embeddings of the Reissner-Nordstr\"om black hole in string theory. We completely characterize the black hole thermodynamics computing the Hawking temperature, Wald entropy, mass, gauge charges and their dual thermodynamic potentials. We verify that all these quantities are related by the first law of extended black hole mechanics and the Smarr formula once we include a potential associated to the dimensionful parameter $\alpha'$ and the scalar charges. We found that the latter are not identified with the poles at infinity of the scalar fields because they receive $\alpha'$ corrections. 
		\end{minipage}
	\end{center}
	\newpage
	\setcounter{page}{1}
	\pagestyle{plain}
	\renewcommand{\thefootnote}{\arabic{footnote}}
	\setcounter{footnote}{0}
	
	
	\tableofcontents
	
	
\section{Introduction}
Black hole (BH) solutions are an excellent setup to test string theory as a theory of quantum gravity. The matching between the Bekenstein-Hawking entropy and the microscopic states counting for the 5-dimensional BPS black hole solution of the string effective action considered by Strominger and Vafa \cite{Strominger:1996sh} is still one of the main successes of Superstring theory. Soon after, the non-extremal version of such BH has been studied and using U-duality arguments some limits of the Bekenstein-Hawking entropy have been matched with the microscopic states counting \cite{Horowitz_1996}. Since then, a lot of work has been done to extend such results to other solutions and at higher order in $\alpha'$. From the macroscopic side, a major development is due to the introduction of the entropy function formalism \cite{Sen:2007qy}, whose ideas have been used to compute the corrections to the entropy of asymptotically-flat extremal BHs using only their near horizon limit \cite{Castro:2007hc, Castro:2007ci, Castro:2008ne, DominisPrester:2008ynb}. From the microscopic side, the corrections to the Cardy formula have been computed in \cite{Castro:2008ys, Kutasov:1998zh, Kraus:2005vz} finding a perfect match.

Despite this success, a proof of the existence of a regular BH connecting the computed near horizon metric and an asymptotically flat region was lacking. A fully analytical extremal supersymmetric solution with near horizon metric $AdS_2 \times S^3$ has been found for the first time in \cite{Cano:2018qev}, allowing to evaluate independently the corrections to the thermodynamic quantities which require the knowledge of the asymptotic fall-off. In a similar fashion, the $\alpha'$ corrections to more general families of charged static extremal BH solutions and stationary BH solutions were computed \cite{Chimento:2018kop,Cano:2018brq, Cano:2019ycn, Cano:2021rey, Cano:2021nzo, Ortin:2021win} increasing the landscape of the corrections already known \cite{CAMPBELL1992199,Natsuume:1994hd,Giveon:2009da}.
Only very recently the corrections to the non-extremal 3-charge, 5-dimensional BH of Strominger and Vafa have been obtained \cite{Cano:2022tmn}. Another recent development in the computation of $\alpha'$ corrections to the thermodynamics is the method described in \cite{Reall:2019sah} to determine higher derivative corrections. The method essentially allows to compute the first order higher derivative corrections to the thermodynamic using the knowledge of the zeroth order solution only. Such advancement makes no longer necessary to solve the corrected equations of motion (EOMs), but it does not spoil the relevance of obtaining an analytical correction. Indeed, the extra information contained in the analytical solution at first order in $\alpha'$ is still relevant because it can be used to obtain the second order corrections as described in \cite{Ma:2023qqj} and explicitly applied in \cite{Cano:2023dyg}. Moreover, the method of \cite{Reall:2019sah} has never been applied and tested with 10-dimensional Heterotic String Theory (HST) effective action at first order in $\alpha'$. 

The goal of this paper is to improve our understanding of the picture determining analytically the corrections to the non-extremal version of the 4-charge, 4-dimensional families of BH solutions considered in \cite{Johnson:1996ga,Maldacena:1996gb,Horowitz:1996ac}. The corrections of such solutions in the extremal case have been computed in \cite{Cano:2021nzo}. We do not address the most general case, i.e. with four independent charges, but we focus on the simpler case in which we take 2 of the 4 charges equal and we have only 3 independent charges. Moreover, we aim to give a complete description of the solution thermodynamics.

The setup we are working with is that of the 10-dimensional HST effective action in the Bergshoeff-de Roo formulation \cite{Bergshoeff:1989de}, with fermions and Yang-Mills fields consistently truncated. Among the string theory effective theories, HST has two properties which make it special and particularly suitable to obtain explicit solutions. On one hand it is the only 10-dimensional effective action which has $\alpha'$ corrections already at first order. On the other hand, most of the higher derivative contributions to the EOMs are proportional to the zeroth order EOMs.\footnote{See the Lemma proven in \cite{Bergshoeff:1989de}.} Therefore, the first  order EOMs take a much simpler form when evaluated for a correction of a solution of the zeroth order EOMs. On top of that, we have full control of the supersymmetry transformations.

In order to solve the 10-dimensional EOMs we start by making a spherically symmetric ansatz suited to perform a dimensional reduction over a $\text{T}^6$ torus. Knowing a priori the number of independent unknown functions we need is not a simple task. However, following the logic of \cite{Cano:2021nzo}, we can obtain constraints among them using the duality transformations of the Heterotic effective action. With such constraints we are able to solve some of the EOMs with standard methods. Some of them, however, reduce to higher order differential equations once combined and which we can solve only with the help of a symbolic manipulation program using the technique of \cite{Cano:2022tmn}. 

The duality transformation we use is T-duality. T-duality arises because of the presence of toroidal compact directions, and it takes its simplest form when expressed in term of the fields obtained performing the dimensional reduction. Such representation \cite{Bergshoeff:1994dg} is equivalent to the well-known Buscher rules \cite{BUSCHER198759,BUSCHER1988466} and has been used to extend them to type II theories \cite{Bergshoeff:1995as,Meessen:1998qm} and in HST effective action at first order in $\alpha'$ \cite{Bergshoeff:1995cg,Elgood:2020xwu}.
The reason why we can use T-duality transformations to constrain the unknown functions of the ansatz is that all the lower dimensional fields descending from the dimensional reduction of the Kalb-Ramond (KR) field receive explicit $\alpha'$ corrections. These explicit corrections are interchanged and mixed with the implicit corrections contained in the unknown functions. The explicit corrections can be evaluated exactly because they only require the knowledge of the zeroth order solutions. They can then be used to constrain non-trivially the implicit corrections of the unknown functions.  


In order to compute the corrections to the macroscopic entropy we compute the Wald entropy. However, we can not directly apply Iyer and Wald's entropy formula \cite{Wald0ab,Wald:1993nt, Iyer:1994ys} because of the presence of Chern-Simons terms in the KR field strength, as it is well understood \cite{Elgood:2020nls,Cano:2022tmn}. A first strategy to deal with Chern-Simons terms was proposed in \cite{Sahoo:2006pm} and successfully applied in \cite{DominisPrester:2008ynb,Faedo:2019xii}. Recently, an extension of Wald algorithm has been proposed \cite{Elgood:2020svt,Elgood:2020mdx,Elgood:2020nls} in order to obtain an entropy formula explicitly gauge invariant and frame independent. The entropy formula proposed in \cite{Elgood:2020nls} has been successfully tested in examples \cite{Cano:2019ycn,Cano:2022tmn} and it is the same we use in the current work. 

The program of revisiting the Wald formalism which started with the research of a gauge invariant and frame independent entropy formula has recently developed further \cite{Ortin:2021ade,Mitsios:2021zrn,Meessen:2022hcg,Ortin:2022uxa,Ballesteros:2023iqb,Gomez-Fayren:2023wxk,Bandos:2023zbs,Ballestaros:2023ipa}. The main advancements are related to the understanding of the role of the chemical potentials associated to the magnetic charges \cite{Ortin:2022uxa}, the role of the chemical potentials associated to the dimensionful parameters appearing in the effective action \cite{Meessen:2022hcg} (see \cite{Kastor:2009wy} for the seminal work on this topic) and the role of the scalar charges \cite{Ballesteros:2023iqb} (first studied in \cite{Gibbons:1996af}). These works produced proposals not specific for HST which can be tested with our analytical solution. For instance, an highly non-trivial test for the entropy formula proposed is the matching between the Hawking temperature $T_H$ and the temperature obtained from the thermodynamic relation $\delta S/\delta M = 1/T$.
  
This paper is organized as follows: in section \ref{sec01} we present the 10-dimensional ansatz we use and the induced 4-dimensional fields. In section \ref{secResEQs} we describe the steps we followed to solve the EOMs of HST and to fix the integration constants imposing the regularity of the solution of the BH horizon. Here we give the explicit analytical expression of the corrections. In section \ref{secTherm} we discuss the thermodynamics of the solution. In particular, we compute the gauge charges, the mass, the Hawking temperature, the Wald entropy and the chemical potentials associated to the gauge charges. All these quantities satisfy the first law of black hole mechanics if keep the moduli fixed. The Smarr formula is satisfied provided that we introduce a proper potential associated to the dimensionful parameter $\alpha'$. Allowing the values of the moduli to vary, we determine the scalar charges. We found that in the case of the dilaton, the scalar charge is not identified with the pole at infinity and it receives $\alpha'$ corrections. In section \ref{secConclusions} we discuss the results and possible future developments. In appendix \ref{sec-heteroticalpha} we review the HST effective action and our conventions. In appendix \ref{sec10to4d} we report the dictionary between higher and lower dimensional fields in toric dimensional reductions. In appendices  \ref{secTdual} and \ref{secChargesandMass} we derive relations used in the main text. In the first one, we derive the constraints imposed by T-duality on the ansatz we made. In the second one, we derive the constraints we get on the integration constants once we relate them to the gauge charges and the mass. In section \ref{secLimits} we report some relevant limits of the solution and the thermodynamics we obtained. We briefly compare the results with the previous literature.

\section{The Ansatz} \label{sec01}

Some non-extremal BH solutions can be obtained simply adding the so called \textit{blackening factor} to the ansatz of the extremal solutions (see for instance \cite{Ortin:2015hya} for a review). This is in particular the case for the BH solutions of heterotic string theory at zeroth order in $\alpha'$ we considered in \cite{Cano:2021nzo}, representing 3-charged extremal BHs in 5 dimensions and 4-charged extremal BHs in 4 dimensions. In \cite{Cano:2022tmn} we verified that such possibility is not spoiled by $\alpha'$ corrections for the families of 3-charged extremal BH solutions of \cite{Cano:2021nzo}. A natural guess to obtain a non-extremal 4-charged black hole is, then, the ansatz for a 4-charge, extremal BH dressed with blackening factors.

\subsection{10d form}

We consider the 10d ansatz for the metric, the KR field strength and the dilaton 
\begin{subequations}\label{eq021}
\begin{align}
	\begin{split}
		d\hat{s}^2 \, = & \;\; \frac{W_{tt}}{\mathcal{Z}_+\mathcal{Z}_-}dt^2-\mathcal{Z}_0\mathcal{Z}_\mathcal{H}(W_{rr}^{-1} dr^2 + r^2d\Omega_{(2)}^2) \\[2mm] & \, -\ell_{\infty }^2\frac{\mathcal{Z}_0}{\mathcal{Z}_\mathcal{H}} \bigg[dw + \ell_{\infty }^{-1}\beta_{\mathcal{H}} \, q_{\mathcal{H}} \cos\theta d\varphi \bigg]^2 \\[2mm] & \, -k_{\infty}^2\frac{\mathcal{Z}_+}{\mathcal{Z}_-} \left[\,dz + k_{\infty}^{-1}\beta_+(\mathcal{Z}^{-1}_+-1)\,dt\right]^2 - dy^{\tilde{m}} dy^{\tilde{m}}, \quad \quad \tilde{m}=1,\dots 4\,,
	\end{split} \\[2mm]
	\hat{H} \, = & \;\; k_{\infty}\beta_- \,d\left[(\mathcal{Z}^{-1}_{h-}-1)\,dt\wedge dz\right] + \ell_{\infty }\beta_{0} \, r^2 \mathcal{Z}_{h0}'\,\omega_{(2)} \wedge d w\,, \\[2mm]
	e^{-2\hat{\phi}} \, = & \; \; -\frac{c_{\hat{ \phi}}}{r^2 \mathcal{Z}_{h-}'} \sqrt{\frac{W_{tt}}{W_{rr}}} \left(\frac{\mathcal{Z}_{h-}}{\mathcal{Z}_{-}}\right)^2 \frac{\mathcal{Z}_-}{\mathcal{Z}_0}\,, 
\end{align}
\end{subequations}
with
\begin{subequations}
	\begin{align}
	d\Omega_{(2)}^2 & = d\theta^2+\sin(\theta)^2 d\varphi^2 \,, \\[2mm]
	\omega_{(2)} & = \sin \theta \, d\theta \wedge d\varphi \,.
	\end{align}
\end{subequations}
The coordinates are adapted to perform a dimensional reduction over the internal manifold $\mathcal{M}_6 = \text{S}^1 \times \text{S}^1 \times \text{T}^4$. The internal circles are parameterized by $z$ and $w$ which satisfy the periodicity conditions
\begin{equation}
	z \sim z + 2 \pi \ell_s \,,  \qquad	w \sim w + 2 \pi \ell_s \,,
\end{equation} 
where $\ell_s = \sqrt{\alpha'}$ is the string length. $k_{\infty}$ and $\ell_{\infty }$ are moduli corresponding to the  asymptotic values of the scalar fields describing the radii of the internal circles in string units 
\begin{equation}
	\text{vol}(\text{S}^1_{\infty,z})/2\pi = R_z \equiv k_{\infty} \ell_s \,, \qquad 	\text{vol}(\text{S}^1_{\infty,w})/2\pi = R_w \equiv \ell_{\infty} \ell_s \,.
\end{equation}
$c_{\hat{ \phi}}$ is a constant related to the asymptotic value of the dilaton that we will fix later. The ansatz for the dilaton has been chosen in such a way that the Kalb-Ramond (KB) EOM is automatically satisfied. Notice that we have assumed that the asymptotic value of the scalars associated to the $\text{T}^4$ is 1, in such a way that $\text{vol}(\text{T}^4) = (2\pi \ell_s)^4 $.

In order to describe a 4-charge configuration, our ansatz for the unknown functions is
\begin{equation}
	\mathcal{Z}_i = 1 + \frac{q_i}{r} + \alpha' \delta \mathcal{Z}_i \,, \qquad \mathcal{Z}_{hi} = 1 + \frac{q_i}{r} + \alpha' \delta \mathcal{Z}_{hi} \,, \qquad 
	 W_{j} = 1 + \frac{\omega}{r} +\alpha' \delta {W}_j \,,
\end{equation}
where $i \in \{\pm, 0, \mathcal{H}\}$, $hi \in \{ h-,h0\}$ and $j \in \{ tt,rr \}$. The ansatz solves the EOMs at zeroth order if
\begin{equation}
	\omega = q_i \left(1-\beta_i^2\right) \,.
\end{equation} 
We assume an ansatz for the $\beta$s such that this relation is not modified at first order in $\alpha'$, i.e. we consider
\begin{equation}
	\beta_i = s_i\sqrt{1-\frac{\omega}{q_i}}\,,
\end{equation}
where, $s_i$ are signs. In the next sections we will solve the EOMs in the case with 3 independent charges only 
\begin{equation}
	q_{\mathcal{H}} = q_0 \equiv q \,.
\end{equation}
Finally, we assume that  $q_i > 0$ and $\omega < 0$. The former condition is necessary to obtain regular solutions at zeroth order. The second condition can always be satisfied. Indeed, a solution with $\omega > 0$ can be mapped into a solution with $\tilde{\omega} < 0 $ with the change of coordinates and the dictionary between the $q$s\footnote{The statement is easy to verify at zeroth order. We verified it at first order in $\alpha'$ building solutions for both cases and checking that they are mapped into each other.}
\begin{equation}
	\tilde{r} = r + \omega\,, \qquad \tilde{\omega} = - \omega \,, \qquad \tilde{q}_i = q_i - \omega \,.
\end{equation}

\subsection{4d form}

Using the relation between 10d and 4d fields summarized in appendix \ref{sec10to4d}, our ansatz in the 4d string frame takes the form\footnote{In order to perform the dimensional reduction we used the zeroth order solution to verify that $\omega^{(L)}_{trz}$ vanishes. No further on-shell relations have been used. To get an explicit expression for $\hat{B}_{\hat{\mu}\hat{\nu}}$ we fixed some integration constants imposing the absence of $\alpha'$ corrections to the asymptotic value of the fields and to the charge associated to $C_w^{(1)}$. See appendix \ref{secChargesandMass} for further details.} (we omit the indexes over the trivial $\text{T}^4$)
\begin{subequations} \label{ansatz4d}
	\begin{align}
		ds^2 & =
		\frac{W_{tt}}{\mathcal{Z}_{+}\mathcal{Z}_{-}}dt^{2}
		-\mathcal{Z}_{0}\mathcal{H}\left({W}^{-1}_{rr}dr^{2}+r^{2}d\Omega^{2}_{(2)}\right)\,, \\[4mm]
		G_{mn} & \equiv \begin{pmatrix} \ell^2 & 0 \\ 0 & k^2   \end{pmatrix} = \begin{pmatrix} \ell_\infty^2{\mathcal{Z}_0}/{\mathcal{Z}_\mathcal{H}} & 0 \\ 0 & k_\infty^2 {\mathcal{Z}_+}/{\mathcal{Z}_-}  \end{pmatrix} \,, \label{eqscalars4d} \qquad m,n \in \{w, z\} \,, \\[4mm]
		 A^m & = \begin{pmatrix}\ell_\infty^{-1}\beta_\mathcal{H} \, q \cos \theta \, d\varphi 		
		,& k_\infty^{-1}\beta_+\left[-1+\mathcal{Z}_+^{-1}\right]dt  
		\end{pmatrix}\,, \label{eqvecA4d} \\[4mm]
		 C^{(1)}_m & = \begin{pmatrix}
		\ell_\infty \beta_0 \, q_0 \cos \theta \, d\varphi  , & 	k_\infty \beta_-\left[-1+\mathcal{Z}_{h-}^{-1}  \left(1+\alpha' \beta_-^{-1} \Delta_C \right)\right] dt
		\end{pmatrix} \,,  \label{eqvecC4d} \\[4mm] 
		 e^{-2\phi} & =  -\frac{c_{{ \phi}}}{r^2 \mathcal{Z}_{h-}'} \sqrt{\frac{W_{tt}}{W_{rr}}} \left(\frac{\mathcal{Z}_{h-}}{\mathcal{Z}_{-}}\right)^2 \sqrt{\frac{\mathcal{Z}_+\mathcal{Z}_-}{\mathcal{Z}_\mathcal{H}\mathcal{Z}_0}}  \label{eqdilaton4d} \,,
	\end{align}
\end{subequations}
with $c_{\phi} = c_{\hat{\phi}} k_\infty l_\infty$ and
\begin{equation}
	\Delta_C = \frac{-W'\left(\beta_{-}\mathcal{Z}_+\mathcal{Z}_{-}'+\beta_+\mathcal{Z}_+'\mathcal{Z}_-\right)+2\left(\beta_{+}+\beta_{-}\right)\mathcal{Z}_+'W\mathcal{Z}_-'}{8\,\mathcal{Z}_0\mathcal{Z}_-\mathcal{Z}_+} + \mathcal{O}(\alpha) \,.
\end{equation}
The metric in the \textit{modified Einstein frame}\footnote{It is the unique Einstein frame in which the metric is asymptotically flat with the standard normalization. It has been introduced in \cite{Maldacena:1996ky}} takes the form
\begin{equation}
	ds^2_E = ds^2 e^{-2(\phi-\phi_\infty)} = F \left[\frac{W_{tt}}{f} dt^2 - f\left(W_{rr}^{-1}dr^2 + r^2 d\Omega_{(2)} \right)\right] \,,
\end{equation}
where 
\begin{equation}
	F = -\frac{c_{{ \phi}}\,e^{2\phi_\infty}}{r^2 \mathcal{Z}_{h-}'} \sqrt{\frac{W_{tt}}{W_{rr}}} \left(\frac{\mathcal{Z}_{h-}}{\mathcal{Z}_{-}}\right)^2 \,, \qquad f = \sqrt{\mathcal{Z}_+\mathcal{Z}_-\mathcal{Z}_0\mathcal{Z}_\mathcal{H}} \,.
\end{equation}
For the other 4d fields we introduce instead the \textit{modified Einstein normalization}. It is the global rescaling of the 4d fields that absorb all the explicit occurrence of the moduli in the action.\footnote{Such normalization has been introduced in \cite{Gomez-Fayren:2023wxk} without a specific name.} The fields in the modified Einstein normalization take the form
\begin{subequations}
\begin{align}
	G_{mn}{}_E & = G_{mn} \,, \\[2mm]
	 A^m_E & = A^m e^{\phi_\infty} \,, \\[2mm]
	  C^{(1)}_m{}_E & =  C^{(1)}_m  e^{\phi_\infty} \,, \\[2mm]
	   e^{-2\phi_E} & =  e^{-2\phi} \,.
\end{align}
\end{subequations}
The modified Einstein normalization will have a central role in the study of the BH thermodynamics. In order to not treat the metric separately from the other fields, in the rest of the paper we will refer to the metric in the modified Einstein frame as the metric in the modified Einstein normalization.

Finally, the combinations $k^{(1)}$ and $\ell^{(1)}$ defined in appendix \ref{sec10to4d} and involved in T-duality transformations (see appendix \ref{secTdual})  take the form
\begin{equation}
	 \qquad k^{(1)}  =  k_\infty\sqrt{\frac{\mathcal{Z}_+}{\mathcal{Z}_-}} \left(1+\alpha' \Delta_k\right)\,,	 
	 \qquad \ell^{(1)}  = \ell_\infty \sqrt{\frac{\mathcal{Z}_0}{\mathcal{Z}_\mathcal{H}}}\left(1+\alpha' \Delta_\ell\right)\,,
\end{equation}
with
\begin{subequations}
	\begin{align}
		& \Delta_k = \frac{-W\left(\mathcal{Z}_+\mathcal{Z}_-'-\mathcal{Z}_-\mathcal{Z}_+'\right)^2+\left(\beta_{-}\mathcal{Z}_+\mathcal{Z}_{-}'+\beta_+\mathcal{Z}_{-}\mathcal{Z}_+'\right)^2}{8\,\mathcal{Z}_0\mathcal{Z}_-^2\mathcal{Z}_+^2} + \mathcal{O}(\alpha)\,, \\[2mm]
		& \Delta_\ell = \frac{-W\left(\mathcal{Z}_0\mathcal{Z}_\mathcal{H}'-\mathcal{Z}_\mathcal{H}\mathcal{Z}_0'\right)^2-\left(\beta_\mathcal{H} \mathcal{Z}_0\mathcal{Z}_\mathcal{H}'+\beta_0\mathcal{Z}_\mathcal{H}\mathcal{Z}_0'\right)^2}{8\,\mathcal{Z}_0^3 \mathcal{Z}_\mathcal{H}^3} + \mathcal{O}(\alpha) \,.
	\end{align}
\end{subequations}

\section{Black Hole Solutions} \label{secResEQs}

In this section we describe the steps we followed to solve the EOMs and determine the integration constants. We conclude presenting the explicit analytical solutions.    

\subsection{Solving the EOMs}

Replacing the ansatz (\ref{eq021}) into the EOMs and the Bianchi identity of HST (see appendix \ref{sec-heteroticalpha} for a review) we can easily fix some of the unknown functions. The KR equation is automatically satisfied and has been used to determine the expression of the dilaton $\hat{\phi}$. Expanding the Bianchi identity in $\alpha'$ and dropping $\mathcal{O}(\alpha'{}^2)$ terms we obtain a second order differential equation for $\delta \mathcal{Z}_{h0}$. It can be solved easily providing
\begin{equation}
	\begin{split}
	\delta \mathcal{Z}_{h0} = & (1+s_0 s_{\mathcal{H}})\left[\frac{q^4+q^3 \omega+11 q^2 r \omega+15 q r^2 \omega+6 r^3 \omega}{4 q^2 r (q+r)^3}-  \frac{3 \omega }{2q^3}\log \left(\frac{q+r}{r}\right)\right]+ \frac{d_{h0}^{(1)}}{r}\,,
	\end{split}
\end{equation}
where $d_{h0}^{(1)}$ is an integration constant (the second integration constant has been already fixed asking that the asymptotic value of $\mathcal{Z}_{h0}$ is not modified). Imposing that the charge associated with $C^{(1)}_w$ is not renormalized we obtain $d_{h0}^{(1)} = 0$ (see appendix \ref{secChargesandMass}). The dilaton and Einstein equations form instead a complicate system of coupled differential equations. Once we replace our ansatz together with the expressions for $\hat{\phi}$ and $\delta\mathcal{Z}_{h0}$ and we drop $\mathcal{O}(\alpha'{}^2)$ terms we obtain a total of 9 non trivial equations (we indicate with $\mathbb{E}_{\hat{\mu}\hat{\nu}}$ the components of the Einstein equations (\ref{eq:eq1}) and with $\mathbb{E}_\phi$ the dilaton EOM (\ref{eq:eq2}))
\begin{equation}
	\{\mathbb{E}_{tt}, \mathbb{E}_{rr},\mathbb{E}_{\theta\theta},\mathbb{E}_{\varphi\varphi},\mathbb{E}_{ww},\mathbb{E}_{zz},\mathbb{E}_{tz},\mathbb{E}_{\varphi w},\mathbb{E}_{\phi}\}\,.
\end{equation}
However, not all of them are independent. We can drop for instance $\mathbb{E}_{\varphi\varphi}$ and $\mathbb{E}_{\varphi w}$ because they are combinations of the other EOMs. $\mathbb{E}_{ww}$ turns out to be a second order differential equation for the combination $\delta \mathcal{Z}_{\mathcal{H}} - \delta \mathcal{Z}_0$. Solving it we obtain
\begin{equation}
	\begin{split}
	\delta \mathcal{Z}_{\mathcal{H}}  = &  \; \delta \mathcal{Z}_0  + (d_0^{(1)}-d_\mathcal{H}^{(1)})\left[\frac{4 q (q-\omega)}{r \omega^2}- \frac{(2 q - \omega)(2 q r + q w - r \omega)}{r \omega^3}\log\left(1+\frac{\omega}{r}\right)\right] \\& 
	- \frac{q(1+s_0s_{\mathcal{H}})(q-\omega)}{4 r (q + r)^3} \,,
\end{split}
\end{equation} 
where we imposed that both $\mathcal{Z}_{\mathcal{H}}$ and $\mathcal{Z}_{0}$ vanish asymptotically. $d_0^{(1)}$ and $d_\mathcal{H}^{(1)}$ are integration constants and represent the poles of the $1/r$ terms of $\delta \mathcal{Z}_{0} $ and $\delta \mathcal{Z}_{\mathcal{H}} $. $\mathbb{E}_{\theta\theta}$ is instead an algebraic constraint for $\delta W_{rr}$. We can use it to fix  $\delta W_{rr}$ as a function of the other unknown functions an their derivatives (we omit at this stage the actual expression because of its lengthiness)
\begin{equation}
	\delta W_{rr} = f \left(r,\delta W_{tt},\delta W_{tt}',\delta\mathcal{Z}_0,\delta\mathcal{Z}_0',\delta\mathcal{Z}_0'',\delta\mathcal{Z}_-,\delta\mathcal{Z}_-',\delta\mathcal{Z}_{h-},\delta\mathcal{Z}_{h-}',\delta\mathcal{Z}_{h-}''\right)\,.
\end{equation}
We are left with 5 equations and 5 unknown functions. Despite the complexity of the system is possible to solve it with the same procedure of \cite{Cano:2022tmn}. First, we consider an ansatz for the unknown functions with arbitrary coefficients
\begin{equation}
	\delta \mathcal{Z}_i = \sum_{k>0} \frac{d_i^{(k)}}{r^k} \,, \qquad \delta \mathcal{Z}_{h-} = \sum_{k>0} \frac{d_{h-}^{(k)}}{r^k} \,, \qquad
	\delta W_{j} = \sum_{k>0} \frac{d_{wj}^{(k)}}{r^k}\,,
\end{equation}
where we assumed only that $\alpha'$ corrections do not modify the asymptotic value of the $\mathcal{Z}$s and $W$s functions. Then, we replace the series expansions into the EOMs and we demand that the they are solved order by order in powers of of $1/r$. In this way we obtain a set of algebraic equations for the coefficients $d^{(k)}$ for arbitrarily large values of  $k$. Solving such equations we obtain the asymptotic expansion in powers of $1/r$ of the unknown functions. The coefficients of one of these functions are not fixed, signaling that only 4 of the 5 EOMs left are truly independent. More precisely, we find a family of solutions which depend on the coefficients
\begin{equation}
	\left\{d_{h-}^{(1)}, d_-^{(1)},d_+^{(1)},d_\mathcal{H}^{(1)},d_{0}^{(1)},d_{wt}^{(1)},d_{wr}^{(1)}, d_{h-}^{(k)}  \right\} \,, \qquad k \ge 3 \,.
\end{equation}
Imposing that the charges associated with $A^z$ and $C_z^{(1)}$ are not renormalized (i.e. they do not receive $\alpha'$ corrections) we fix $d_{h-}^{(1)} = d_+^{(1)} = 0$ (see appendix \ref{secChargesandMass}). We then set to zero the coefficients $d_{h-}^{(k)}$ with $k\ge3$. This can always be done without loss of generality, because it is equivalent to perform a proper change of coordinates. We obtain the simple expression for $\delta \mathcal{Z}_{h-}$
\begin{equation}
	\delta \mathcal{Z}_{h-} = \frac{1}{2}\left(d_{wt}^{(1)}-d_{wr}^{(1)}-2d_{-}^{(1)}\right)\frac{q_-}{r^2} \,.
\end{equation}
With this expression of  $\delta \mathcal{Z}_{h-}$ we can easily determine some other quantities. First of all we notice that expanding the dilaton in series we get
\begin{equation}
	e^{-2\phi} \sim \frac{c_\phi}{q_-} + \mathcal{O}(1/r) \,,
\end{equation}
which allows us to fix $c_\phi = e^{-2\phi_\infty} q_-$. Second, we can use the compatibility with the T-duality constraints to extract $\mathcal{Z}_{\pm}$ (see appendix \ref{secTdual}). The action of T-duality along the $z$ direction on the lower dimensional fields is 
\begin{equation}
	T_z: \qquad 	C^{(1)}_z \leftrightarrow A^z \,, \qquad k  \leftrightarrow 1/k^{(1)} \,, \qquad	ds^2_{E} \leftrightarrow ds^2_{E} \,, \qquad e^{-2\phi}\leftrightarrow e^{-2\phi} \,. \\
\end{equation}
Assuming that $T_z$ can be implemented by
\begin{equation} 
		T_z: \qquad 	q_+ \leftrightarrow q_- \,, \qquad \beta_+ \leftrightarrow \beta_- \,, \qquad k_\infty \leftrightarrow 1/k_\infty \,,
\end{equation}
we obtain relation (\ref{eqtdualzmbis}) which provides
\begin{equation}\label{eqdeltazm}
	\delta \mathcal{Z}_- = \delta \mathcal{Z}_{h-} + \mathcal{Z}_- \left[\Delta_k - \frac{\Delta_C}{\beta_+}+\frac{r^2}{2}\left(\frac{\delta \mathcal{Z}_{h-}'}{q_-} - \frac{T_z[\delta\mathcal{Z}_{h-}']}{q_+}\right)\right] \,,
\end{equation} 
where $T_z$ is the operator implementing the T-duality transformation. Such assumption imposes non-trivial constraints on some integration constants appearing in the series describing the expansion of $\delta \mathcal{Z}_-$. Indeed, the series $\{d^{(k)}_-\}$ satisfies (\ref{eqdeltazm}) provided that
\begin{equation}
	T_z \left[d_{h-}^{(2)}\right] = \frac{1}{2} q_+ \left(d_{wt}^{(1)}- d_{wr}^{(1)}\right) \,, \qquad d_-^{(1)} = \frac{(q_+-q_-)}{2q_- - w}d_{wt}^{(1)} \,.
\end{equation}
Expanding equation (\ref{eqtdualzmh}) we obtain an expression for $\delta \mathcal{Z}_+$ which matches the series $\{d_+^{(k)}\}$ without further constraints
\begin{equation}
	\delta \mathcal{Z}_+ = T_z \left[\delta \mathcal{Z}_{h-}\right] - \mathcal{Z}_+ \frac{\Delta_C}{\beta_+}\,.
\end{equation}
Once we replace the expressions obtained for $\delta \mathcal{Z}_-$, $\delta \mathcal{Z}_+$, $\delta\mathcal{Z}_\mathcal{H}$, $\delta\mathcal{Z}_{h-}$ and $\delta W_{rr}$ into the EOMs we obtain a set of differential equations for $\delta W_{tt}$ and $\delta \mathcal{Z}_0$. In particular, $\mathbb{E}_{zz}$ is a first order differential equations which involves only $\delta W_{tt}$. Solving it we obtain
\begin{equation}
	\begin{split}
	\delta W_{tt} = \quad &\frac{ r \omega^2-q_- \omega (2 r+\omega)}{8 r^2 (q+r)^2 (q_-+r)}-\frac{\beta_- \beta_+ q_- q_+ \omega (r+\omega)}{4 r^2 (q+r)^2 (q_-+r) (q_+-\omega)} \\
	&  + \frac{d_{wt}^{(1)} (2 q_++2 r+\omega)}{2 r^2}-\frac{d_{wr}^{(1)} \omega}{2 r^2} \,.
\end{split}
\end{equation}
Replacing this last expression for $\delta W_{tt}$ into the EOMs, we are left with a fourth order differential equation for $\delta \mathcal{Z}_0$. Instead of solving it directly, we focus on finding the generating function of the coefficients $\{d_0^{(k)}\}$. We actually addressed this step using a symbolic manipulation program that is capable to make a good guess for the generating function given a very large number of terms of the series. Replacing the guessed generating function into the fourth order order differential equation we can verify that it is exactly solved and we actually find the expression of $\delta \mathcal{Z}_0$ (we omit again the actual expression because of its lengthiness). With the explicit expression of $\delta \mathcal{Z}_0$ one can finally reconstruct the explicit expressions of all the $\delta \mathcal{Z}$s and $\delta W$s. It is then possible to verify that the expressions obtained solve exactly all the EOMs of HST.

\subsection{Regularity conditions}

At zeroth order the horizon lies at $r = -\omega$. At first order it may be shifted and placed at $r_H =-\omega + \alpha' \delta r$. Therefore, we expand the 4d fields around $\rho = (r-r_H)$. We obtain
\begin{subequations}
\begin{align}
	 e^{-2 \phi } & \quad  = \quad  y^{(0)}_\phi + y^{(0,\log)}_\phi \log \rho + \mathcal{O}(\rho \log \rho) \,, \\
	k & \quad  = \quad y^{0}_k + \mathcal{O}(\rho) \,, \\
	\ell & \quad  = \quad y^{(0)}_\ell + y^{(0,\log)}_\ell \log \rho + \mathcal{O}(\rho \log \rho)\,, \\
	g_{tt,E} & \quad  = \quad y^{(0)}_{tt} +  y^{(1,\log)}_{tt} \rho \log \rho + \mathcal{O}(\rho) \,, \\
	g_{rr,E} & \quad  = \quad \frac{y^{(-2)}_{rr} }{\rho^2} +\frac{y^{(-1)}_{rr} }{\rho} +\frac{y^{(-1,\log)}_{rr} }{\rho} \log \rho + y^{(0,\log)}_{rr} \log \rho + \mathcal{O}(1) \,, \\
	g_{\theta\theta,E} & \quad  = \quad 	y^{(0)}_{\theta\theta} +  \mathcal{O}(\rho \log \rho) \,, \\
	F^w_{\theta \varphi} & \quad  =  \quad y^{(0)}_{Fw}  \,, \\
	F^z_{tr}  & \quad  = \quad  y^{(0)}_{Fz} + \mathcal{O}(\rho)\,, \\
	G_w & \quad  = \quad y^{(0)}_{Gw}  \,, \\
	G_z & \quad  = \quad   y^{(0)}_{Gz} + \mathcal{O}(\rho) \,,
\end{align}
\end{subequations}
where the $y_i^{(k)}$s are combinations of $q_i$, $\omega$, $\beta_i$, $d_\mathcal{H}^{(1)}$, $d_{0}^{(1)}$, $d_{wt}^{(1)}$,  $d_{wr}^{(1)}$, $\delta r$, $\phi_\infty$, $k_\infty$, $\ell_\infty$  and $\alpha'$. Imposing that the BH horizon is placed at $\rho = 0$ we obtain the condition 
\begin{equation} \label{eqcondreg1}
y^{(0)}_{tt} = 0 \,.
\end{equation}
Imposing that the scalars have a finite value on the BH horizon we get
\begin{equation}\label{eqcondreg2}
	 y^{(0,\log)}_\phi  = y^{(0,\log)}_\ell = 0 \,.
\end{equation}
Demanding that the Hawking temperature is finite we obtain 
\begin{equation}\label{eqcondreg3}
	 y^{(1,\log)}_{tt} = y^{(-2)}_{rr} = y^{(-1,\log)}_{rr} = 0 \,.
\end{equation}
The conditions (\ref{eqcondreg1}),(\ref{eqcondreg2}),(\ref{eqcondreg3}) together with the requirement that the BH horizon is not shifted, i.e. $\delta r = 0$, lead to
\begin{subequations}
\begin{align}
	d_0^{(1)} & = d_\mathcal{H}^{(1)} \,, \\[2mm]
	\begin{split}
	d_{wt}^{(1)} & = \frac{ \omega (\omega-2 q)}{(q-\omega) (\omega-2 q_+)}d_\mathcal{H}^{(1)} +\frac{ s_0 s_\mathcal{H} \, q \, \omega \left(2 q^2-6 q \omega+5 \omega^2\right)}{20 (q-\omega)^4 (\omega-2 q_+)} \\
	& \quad +\frac{\omega \left(-4 q^3+22 q^2 \omega-35 q \omega^2+25 \omega^3\right)}{40 (q-\omega)^4 (\omega-2 q_+)} \,, 
	\end{split} \\[4mm]
	\begin{split}
		d_{wr}^{(1)} & = \frac{ (2 q-\omega)}{q-\omega}d_\mathcal{H}^{(1)}-\frac{s_0 s_{\mathcal{H}} \,q  \left(2 q^2-6 q \omega+5 \omega^2\right)}{20 (q-\omega)^4} \\
		& \quad +\frac{4 q^3-12 q^2 \omega+15 q \omega^2-15 \omega^3}{40 (q-\omega)^4} \,.
	\end{split}
\end{align}
\end{subequations}

\subsection{Complete Solution}

Properly fixing the $\alpha'$ corrections to the mass (see appendix \ref{secChargesandMass}) we obtain the regular solution 
\begin{subequations}
\begin{align}
	\begin{split}
		\delta \mathcal{Z}_{h0} = & \; (1+s_0 s_{\mathcal{H}})\left[\frac{q^4+q^3 \omega+11 q^2 r \omega+15 q r^2 \omega+6 r^3 \omega}{4 q^2 r (q+r)^3}-  \frac{3 \omega }{2q^3}\log \mathcal{Z}_0 \right]\,,
	\end{split} \\[4mm]
	\begin{split}
		\delta \mathcal{Z}_{0} = & \; \frac{d_{\mathcal{H}}^{(1)}}{r}+\frac{\omega^2 (2 q-\omega) (s_0 s_{\mathcal{H}}+4) (\omega  q-3 \omega  r + 2 rq-2 r^2) }{40 q^3 r (q-\omega)^3}\log \mathcal{Z}_0 \\
		& +\frac{1}{120 q^2 r (q+r)^3 (q-\omega)^3} \bigg[-q r^2 (q-3 \omega) \left(5 q^3+57 q \omega^2-26 \omega^3\right) \\
		& +q^2 r \left(-16 q^4+51 q^3 \omega-165 q^2 \omega^2+205 q \omega^3-51 \omega^4\right) \\
		& +2 q^3 (q-\omega) \left(18 q^3-58 q^2 \omega+41 q \omega^2-19 \omega^3\right) \\
		& +24 r^4 \omega^2 (2 q-\omega)  +36 r^3 \omega^2 (2 q-\omega) (q+\omega) \bigg]  \\
		& + \frac{s_0 s_\mathcal{H}}{
			240 q^2 r (q+r)^3 (q-\omega)^3} \bigg[q^3 (q-\omega) \left(48 q^3-128 q^2 \omega+91 q \omega^2-29 \omega^3\right)\\
		&+q r^2 \left(10 q^4-30 q^3 \omega+9 q^2 \omega^2+86 q \omega^3-39 \omega^4\right)\\
		&+2 q^2 r \left(16 q^4-51 q^3 \omega+30 q^2 \omega^2+20 q \omega^3-9 \omega^4\right)\\
		& +12 r^4 \omega^2 (2 q-\omega)+18 r^3 \omega^2 (2 q-\omega) (q+\omega)\bigg] \,,
	\end{split} \\[4mm]
	\begin{split}
		\delta \mathcal{Z}_{\mathcal{H}} = & \; \delta \mathcal{Z}_0 
		- \frac{q(1+s_0s_{\mathcal{H}})(q-\omega)}{4 r (q + r)^3} \,,
	\end{split} \\[4mm]
	\begin{split}
		\delta \mathcal{Z}_{h-} = & \; \frac{ q_- (2 q-\omega) (q_--\omega)}{r^2 (q-\omega) (\omega-2 q_-)}d_{\mathcal{H}}^{(1)}-\frac{q q_- \left(2 q^2-6 q \omega+5 \omega^2\right) (q_--\omega) (s_{0} s_{\mathcal{H}}-1)}{20 r^2 (q-\omega)^4 (\omega-2 q_-)} \\
		& +\frac{q_- \omega^2 \left[q^2+q (q_--3 \omega)+\omega (4 \omega-3 q_-)\right]}{8 r^2 (q-\omega)^4 (\omega-2 q_-)} \,,
	\end{split} \\[4mm]
	\begin{split}
		\delta \mathcal{Z}_- = & \; \delta \mathcal{Z}_{h-} + \mathcal{Z}_- \left[\Delta_k - \frac{\Delta_C}{\beta_+}+\frac{r^2}{2}\left(\frac{\delta \mathcal{Z}_{h-}'}{q_-} - \frac{T_z[\delta\mathcal{Z}_{h-}']}{q_+}\right)\right] \,,
	\end{split} \\[4mm]
	\begin{split}
		\delta \mathcal{Z}_+ = & \; T_z \left[\delta \mathcal{Z}_{h-}\right] - \mathcal{Z}_+ \frac{\Delta_C}{\beta_+} \,,
	\end{split} \\[4mm]
	\begin{split}
		\delta W_{tt} = & \; \frac{q \omega \left(2 q^2-6 q \omega+5 \omega^2\right) (r+\omega) (s_0 s_{\mathcal{H}}-1)}{20 r^2 (q-\omega)^4 (\omega-2 q_+)} -\frac{\omega (2 q-\omega) (r+\omega)}{r^2 (q-\omega) (\omega-2 q_+)}d_{\mathcal{H}}^{(1)} \\
		& \frac{q^2 \omega^2 (2 q_++2 r+\omega)}{8 r^2 (q-\omega)^4 (\omega-2 q_+)}  +\frac{\omega^2}{8 r (q+r)^2 (q_-+r)} \\
		& +\frac{q_- \omega^2 (2 r+\omega)-q_- q_+ \omega [2 r (\beta_- \beta_++1)+2 \beta_- \beta_+ \omega+\omega]}{8 r^2 (q+r)^2 (q_-+r) (q_+-\omega)} \\
		&+\frac{5 \omega^5}{16 r^2 (q-\omega)^4 (\omega-2 q_+)} +\frac{5 q_+ \omega^4}{8 r^2 (q-\omega)^4 (\omega-2 q_+)} \\
		& -\frac{q \omega^3 (4 q_++5 r+3 \omega)}{8 r^2 (q-\omega)^4 (\omega-2 q_+)} +\frac{5 \omega^4}{8 r (q-\omega)^4 (\omega-2 q_+)}+\frac{3 \omega^4}{16 r^2 (q-\omega)^4} \,,
	\end{split} \\[4mm]
	\begin{split}
		\delta W_{rr} = & \; \frac{(2 q-\omega) (r+\omega)}{r^2 (q-\omega)}d_{\mathcal{H}}^{(1)}+\frac{\omega^2 (2 q-\omega) (r+\omega) (2 r+\omega) (s_0 s_{\mathcal{H}}+4)}{20 q^3 r (q-\omega)^3} \log \mathcal{Z}_0 \\
		& + \frac{(s_0 s_{\mathcal{H}}-1)}{120 q^2 r^2 (q+r)^4 (q-\omega)^4}\bigg[-12 r^6 \omega^2 \left(2 q^2-3 q \omega+\omega^2\right) \\
		& -6 q^7 \omega \left(2 q^2-6 q \omega+5 \omega^2\right) \\
		& -6 r^5 \left(2 q^5-6 q^4 \omega+19 q^3 \omega^2-15 q^2 \omega^3-2 q \omega^4+3 \omega^5\right) \\
		& +r^4 \left(-46 q^6+124 q^5 \omega-177 q^4 \omega^2-5 q^3 \omega^3+125 q^2 \omega^4-45 q \omega^5-6 \omega^6\right)\\
		& -q r^3 \left(64 q^6-136 q^5 \omega+42 q^4 \omega^2+221 q^3 \omega^3-167 q^2 \omega^4+15 q \omega^5+21 \omega^6\right) \\
		& +q^2 r^2 \left(-48 q^6+84 q^5 \omega+37 q^4 \omega^2-157 q^3 \omega^3-3 q^2 \omega^4+53 q \omega^5-26 \omega^6\right) \\
		& +q^3 r \left(-12 q^6-12 q^5 \omega+118 q^4 \omega^2-145 q^3 \omega^3+21 q^2 \omega^4+5 q \omega^5-5 \omega^6\right)	\bigg] \\
		& +\frac{\omega^2}{8 q^2 r^2 (q+r)^2 (q-\omega)^4} \bigg[-4 r^4 \left(2 q^2-3 q \omega+\omega^2\right) \\
		& -q r \left(q^4+9 q^2 \omega^2-7 q \omega^3+3 \omega^4\right)+r^3 \left(-11 q^3+3 q^2 \omega+12 q \omega^2-6 \omega^3\right)\\
		& +q^2 \omega \left(-q^3+2 q^2 \omega-4 q \omega^2+\omega^3\right) \\
		& +r^2 \left(-2 q^4-16 q^3 \omega+17 q^2 \omega^2-3 q \omega^3-2 \omega^4\right)\bigg] \,.
	\end{split} 
\end{align}
\end{subequations}
with
\begin{align}
	\begin{split}
	d_{\mathcal{H}}^{(1)} = \; &  \frac{1}{  40 (q-\omega)^3 (2 q-\omega) D}\bigg\{-8 q^3 (s_0 s_{\mathcal{H}}-1) (q (q_-+q_+)+2 q_- q_+) \\
	& +\omega^3 [(q^2 (48 s_0 s_{\mathcal{H}}-98)+q (q_-+q_+) (54 s_0 s_{\mathcal{H}}-59)+4 q_- q_+ (8 s_0 s_{\mathcal{H}}+7)] \\
	& -4 q \omega^2 [q^2 (8 s_0 s_{\mathcal{H}}-13)+2 q (q_-+q_+) (8 s_0 s_{\mathcal{H}}-13)+q_- q_+ (22 s_0 s_{\mathcal{H}}+13)] \\
	& +4 q^2 \omega [2 q^2 (s_0 s_{\mathcal{H}}-1)+q (q_-+q_+) (9 s_0 s_{\mathcal{H}}-14)+2 q_- q_+ (8 s_0 s_{\mathcal{H}}-3)]\\
	& +\omega^4 (-16 s_0 s_{\mathcal{H}} (2 q+q_-+q_+)+72 q+11 (q_-+q_+))+2 \omega^5 (4 s_0 s_{\mathcal{H}}-9)\bigg\} \,,
	\end{split} \\[2mm]
 	D = & \;-2 q (q_-+q_+-\omega)+3 \omega (q_-+q_+)-4 q_- q_+-2 \omega^2 \,, \label{eqdefD}
\end{align}
and $T_m$ is the T-duality operator
\begin{subequations} \label{eqtdualoperator}
	\begin{align}
		&T_z: \qquad 	q_+ \leftrightarrow q_- \,, \qquad \beta_+ \leftrightarrow \beta_- \,, \qquad k_\infty \leftrightarrow 1/k_\infty \,,\\
		&T_w: \qquad 	q\leftrightarrow q \,,  \qquad \beta_{\mathcal{H}} \leftrightarrow \beta_0 \,, \qquad \ell_\infty \leftrightarrow 1/\ell_\infty  \,.
	\end{align}
\end{subequations}

It can be verified that applying (\ref{eqtdualoperator}) to this solution, the ansatz (\ref{ansatz4d}) obeys the transformation rules 
\begin{subequations}
	\begin{align}
		&T_z: \qquad 	C^{(1)}_z \leftrightarrow A^z \,, \qquad k  \leftrightarrow 1/k^{(1)} \,, \qquad	ds^2_{E} \leftrightarrow ds^2_{E} \,, \qquad e^{-2\phi}\leftrightarrow e^{-2\phi} \,, \\
		&T_w: \qquad 	C^{(1)}_w \leftrightarrow A^w \,, \qquad \ell \leftrightarrow 1/\ell^{(1)} \,, \qquad	ds^2_{E} \leftrightarrow ds^2_{E} \,, \qquad e^{-2\phi}\leftrightarrow e^{-2\phi}  \,.
	\end{align}
\end{subequations}

\begin{figure}[h]
	\centering
	\includegraphics[width=0.6\linewidth]{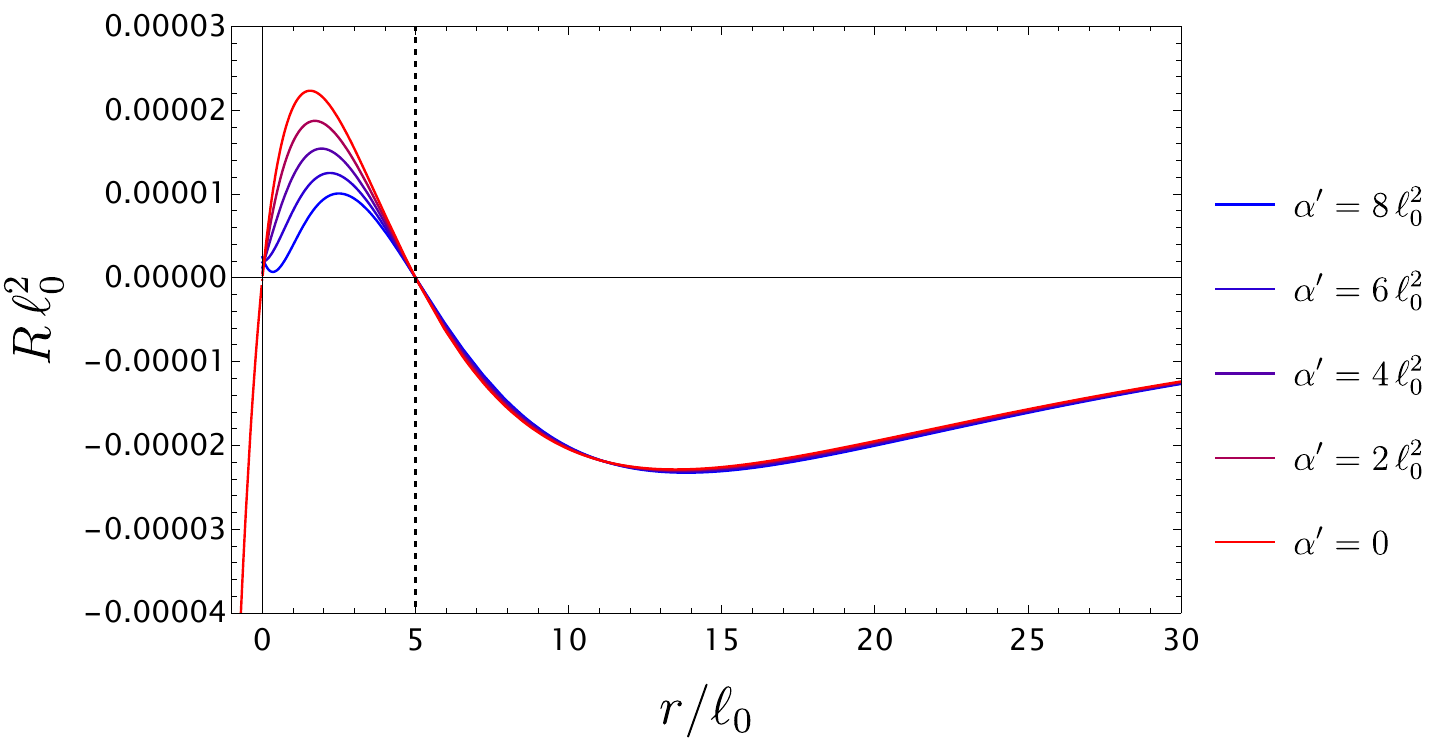}
	\caption{\textit{The Ricci scalar as a function of the radial coordinate for $q_+ = 40 \,\ell_0 $, $q_- = 20 \, \ell_0 $, $q = 10 \, \ell_0 $, $\omega = -5 \, \ell_0 $, $s_+ s_- = s_0 s_{\mathcal{H}} = 1$, for different values of $\alpha'$. We normalized the units setting $\ell_0 = 1$.}}
	\label{figricci}
\end{figure}

\begin{figure}[h]
	\centering
	\includegraphics[width=0.6\linewidth]{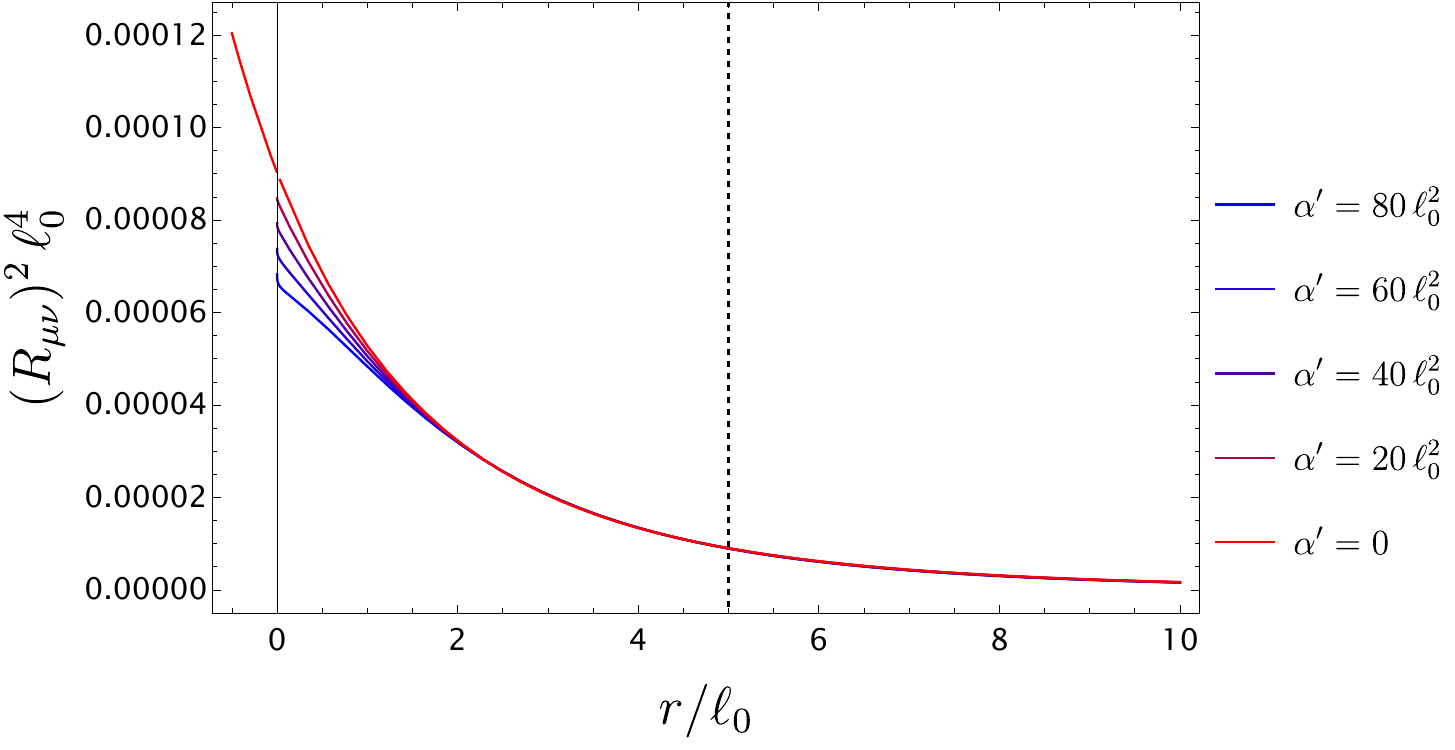}
	\caption{\textit{The $R_{\mu\nu}R^{\mu\nu}$ invariant as a function of the radial coordinate for $q_+ = 40 \,\ell_0 $, $q_- = 20 \, \ell_0 $, $q = 10 \, \ell_0 $, $\omega = -5 \, \ell_0 $, $s_+ s_- = s_0 s_{\mathcal{H}} = 1$ for different values of $\alpha'$. We normalized the units setting $\ell_0 = 1$.}}
	\label{figriccitensor}
\end{figure}

\begin{figure}[h]
	\centering
	\includegraphics[width=0.6\linewidth]{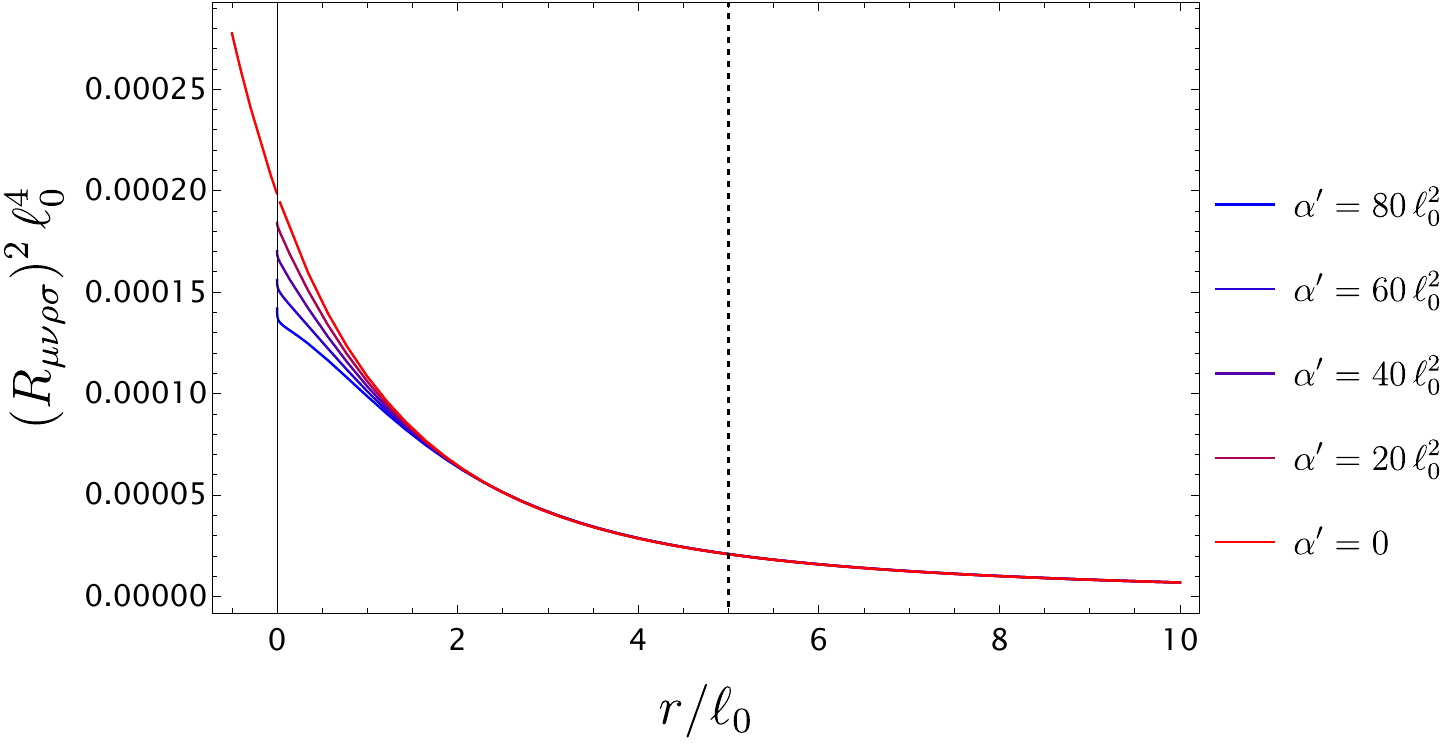}
	\caption{\textit{The Kretschmann invariant $R_{\mu\nu\rho\sigma}R^{\mu\nu\rho\sigma}$ as a function of the radial coordinate for $q_+ = 40 \,\ell_0 $, $q_- = 20 \, \ell_0 $, $q = 10 \, \ell_0 $, $\omega = -5 \, \ell_0 $, $s_+ s_- = s_0 s_{\mathcal{H}} = 1$, for different values of $\alpha'$. We normalized the units setting $\ell_0 = 1$.}}
	\label{figriemann}
\end{figure}

We conclude by plotting some of the curvature invariants (see figures \ref{figricci}, \ref{figriccitensor}, \ref{figriemann}). They show explicitly that the solution obtained for a typical choice of charges and mass within the range of validity of the perturbative regime, does not present curvature singularities outside the BH horizon. The plots do not extend beyond $r = 0$ because there is a logarithmic singularity at that point.


\section{Thermodynamics} \label{secTherm}

In this section we review the thermodynamic properties of our solutions.

\subsection{Gauge Charges and Brane Sources}
Our 4d solution represents a 4-charge BH. More precisely, the BH is electrically charged with respect to two of the gauge vectors, $A^z_E$ and $C^{(1)}_{z\,E}$, and magnetically charged with respect to $A^w_E$ and $C^{(1)}_{w\,E}$.  Explicitly, we have (all the fields are 4d and in the modified Einstein normalization)\footnote{The expression for the electric charge has been obtained integrating the current  $ J = \delta (\star \mathcal{L})/\delta F$ (which is closed on shell) on the surface at infinity and dropping the term which are not contributing.}
\begin{subequations} \label{eqdefCharges}
	\begin{align}
		Q_+ & \equiv \frac{1}{16 \pi G_N^{(4)}} \int_{S^2_\infty} e^{-2\phi}k^2\star_E F^z_E = \frac{k_\infty}{4 G_N^{(4)}g_s^{(4)}} \beta_{+}q_+ \,, \\
		Q_- & \equiv \frac{1}{16 \pi G_N^{(4)}} \int_{S^2_\infty} e^{-2\phi}k^{-2}\star_E G_{z\, E}^{(1)} =  \frac{k_\infty^{-1}}{4 G_N^{(4)} g_s^{(4)}} \beta_{-}q_-  \,, \\
		Q_\mathcal{H} & \equiv -\frac{1}{16 \pi G_N^{(4)}} \int_{S^2_\infty}  F^w_E =  \frac{\ell_\infty^{-1}g_s^{(4)}}{4 G_N^{(4)}} \, \beta_\mathcal{H}\,q  \,, \\
		Q_0 & \equiv -\frac{1}{16 \pi G_N^{(4)}} \int_{S^2_\infty}  G_{w\, E}^{(1)} =  \frac{\ell_\infty g_s^{(4)}}{4 G_N^{(4)}} \, \beta_0\,q \,. 
	\end{align}
\end{subequations}

The normalization of the charges is chosen so that the $Q_i$ have the correct Dirac quantization. With our normalization of the action such condition takes the form (see \cite{Duff:1994an} or \cite{Ortin:2015hya} for a more recent reference)
\begin{equation} \label{eqDirac}
	Q_i Q_j \in \frac{1}{16 \pi G_N^{(4)}} 2 \pi \mathbb{Z} \,.
\end{equation}
One can verify that the charges (\ref{eqdefCharges}) satisfy (\ref{eqDirac}) checking that they are quantized in $1/(\ell_s g_s^{(4)})$ units and recalling that in our conventions 
\begin{equation}
	\left(\ell_s g_s^{(4)}\right)^{-2} = \frac{1}{16\pi G_N^{(4)}} 2\pi \,.
\end{equation}
The quantization of $\mathcal{Q}_\mathcal{H}$ is the simplest to prove. The absence of Dirac-Misner singularities in the 10d ansatz for the metric implies that
\begin{equation}
	N_\mathcal{H} = \frac{2 \, q \beta_{\mathcal{H}}}{R_w}  \in \mathbb{Z} \,,
\end{equation}
and with a straightforward manipulation\footnote{We recall that \begin{equation}
		g_s^{(10-n)}\,{}^2 = g_s^2\, \text{Vol}_n/(2\pi\ell_s)^n \,, \qquad G_N^{(10-n)}  = G_N^{(10)} \, \text{Vol}_n \,,
	\end{equation} with $\text{Vol}(\text{T}^4) = (2\pi \ell_s)^4$,   $\text{Vol}(\text{S}_w^1) = 2\pi\ell_{\infty} \ell_s$,  $\text{Vol}(\text{S}_z^1) = 2\pi k_{\infty} \ell_s$.} we can write the quantized quantity as
\begin{equation}
	N_\mathcal{H}  =  Q_{\mathcal{H}} \ell_s  g_s^{(4)} \,.
\end{equation}
The quantization of $Q_0$ easily follows from the fact that it is the quantity T-dual to $Q_\mathcal{H}$. We have indeed 
\begin{equation}
	Q_\mathcal{H}  =  \frac{\ell_\infty^{-1}g_s^{(4)}}{4 G_N^{(4)}} \, \beta_\mathcal{H}\,q \quad \xrightarrow{T_w} \quad \frac{\ell_\infty g_s^{(4)}}{4 G_N^{(4)}} \, \beta_0\,q = \mathcal{Q}_0 = N_0 \frac{1}{\ell_s  g_s^{(4)}} \,, \qquad N_0 \in \mathbb{Z}\,.
\end{equation}
In order to verify the quantization of $Q_-$ we work in 10d HST. We notice that if we have a total of $N_{-} \in \mathbb{Z}$ fundamental strings wrapped along the $z$ direction and we couple the associated current to the 10d HST effective action, the KR equation takes the form 
\begin{equation}
	\frac{g_s^2}{16 \pi G_N^{(10)}} \int_{V_8} d \left[e^{-2 \hat{\phi}} \hat{\star} \hat{H} + \mathcal{O}(\alpha'{}^2)\right] = T_{F1} N_{-} \,,
\end{equation}
where $V_8$ is such that $\partial{V_8} = \text{T}^4 \times \text{S}^2_\infty \times \text{S}^1_w$ and $T_{F1} = 1/2 \pi \alpha'$. Evaluating the LHS we obtain
\begin{equation}
	\frac{k_\infty^{-1} \ell_s}{4 G_N^{(4)}} \frac{1}{2 \pi \ell_s^2} \beta_- q_- = Q_- g_s^{(4)} \ell_s T_{F1} \equiv T_{F1} N_- \,.
\end{equation}
The quantization of $\mathcal{Q}_+$ follows using T-duality
\begin{equation}
	Q_-  =   \frac{k_\infty^{-1}}{4 G_N^{(4)} g_s^{(4)}} \, \beta_-\,q_- \quad \xrightarrow{T_z} \quad  \frac{k_\infty}{4 G_N^{(4)}g_s^{(4)}} \, \beta_+\,q_+ = \mathcal{Q}_+ = N_+ \frac{1}{\ell_s  g_s^{(4)}}\,, \qquad N_+ \in \mathbb{Z}\,.
\end{equation}

From the point of view of string theory, the solution is a superposition of solitonic 5-branes (NS5) and Kaluza-Klein monopoles (KK6) wrapped around the directions parametrized by the coordinates $y^{1},\cdots,y^{4},z$, fundamental strings (F1) wound around the circle parametrized by $z$ and waves (W) carrying momentum propagating along the same circle (see table \ref{diagram4d}). In the extremal case the integers $N_i$ have a clear interpretation. Their norm is counting the number of sources and their signs distinguish between brane and antibranes. In the non-extremal case the situation is more subtle and the parameters do not have a clear interpretation. However, it has been proposed in similar settings that they may correspond to the difference between the number of branes and the number of antibranes \cite{Horowitz_1996}. 
\begin{table}[h]
	\begin{center}
		\begin{tabular}{c|cccccccccc|c}
			&$t$&$z$&$y^{1}$&$y^{2}$&$y^{3}$&$y^{4}$&$w$&$x^{1}$&$x^{2}$&$x^{3}$ & $\#$\\
			\hline
			F1&$\times$ &$\times$&$\sim$&$\sim$&$\sim$&$\sim$&$\sim$&$-$&$-$&$-$ & $N_-$\\
			\hline
			W&$\times$&$\times$&$\sim$&$\sim$&$\sim$&$\sim$&$\sim$&$-$&$-$&$-$ & $N_+$ \\
			\hline
			NS5&$\times$&$\times$&$\times$&$\times$&$\times$&$\times$&$\sim$&$-$&$-$&$-$ & $N_0$\\
			\hline
			KK6&$\times$&$\times$&$\times$&$\times$&$\times$&$\times$&$-$&$-$&$-$&$-$ & $N_\mathcal{H}$\\
		\end{tabular}
		\caption{\textit{Sources associated to the four-dimensional black holes. The symbol
			$\times$ stands for the worldvolume directions and $-$ for the transverse
			directions. The symbol $\sim$, in turn, denotes a transverse direction over
			which the corresponding object has been smeared.}}
		\label{diagram4d}
	\end{center}
\end{table}

\subsection{Mass, Temperature and Entropy}
The mass of the BH is defined by the asymptotic expansion of the $g_{tt,E}$ component of the metric in the modified Einstein frame
\begin{equation}
	g_{tt,E} \sim 1 - \frac{2 G_N^{(4)} M}{r} + \mathcal{O}(1/r^2) \,.
\end{equation}
We obtain
\begin{equation}
	M = \frac{1}{4 G_N^{(4)}} \left[2 q + q_- + q_+ -2\omega -  (1-s_0 s_\mathcal{H})\frac{\alpha'}{5 (2q-\omega)} \right]\,.
\end{equation}

The Hawking temperature is 
\begin{equation} \label{eqthawking}
 	T_H = \frac{-\omega}{4 \pi\sqrt{ (q-\omega)^2(q_+-\omega)(q_--\omega)}} (1+ \alpha' \Delta_T)\,,
\end{equation}
with
\begin{equation}
\begin{split}
	\Delta_T = & \frac{q_- q_+ (\beta_- \beta_+-1)}{8 (q-\omega)^2 (\omega-q_-) (\omega-q_+)} + \bigg[4 q \omega \left(-5 q^2  -13 qq_--13 qq_++q_- q_+\right) \\
	& +4 q^2 ( 5 qq_-+5 qq_++2 q_- q_+)+\omega^3 (-34 q-8 q_--8q_+) \\
	& +q \omega^2 (50 q+33 q_-+ 33q_+)+8 \omega^4\bigg] \frac{(s_0 s_{\mathcal{H}}-1)}{40 (q-\omega)^3 (2 q-\omega) D} \\
	& + \bigg[\omega^3 \left(-6 q^2 -19 qq_--19 qq_+ +4 q_-^2+26 q_- q_++4 q_+^2\right) \\
	& +\omega^2 \left(12 q^2 q_-+12 q^2 q_+  + 9 qq_-^2+20 q q_- q_++9 q q_+^2-23 q_-^2 q_+ -23 q_- q_+^2 \right)\\
	& +\omega^4 (10 q-4 q_--4 q_+) +\omega \left(20 q_-^2 q_+^2-6 q^2 q_-^2-16 q^2q_- q_+-6 q^2 q_+^2\right)   \\
	& +4 q q_- q_+ (qq_-+qq_+-2 q_- q_+)\bigg] \frac{1}{8 (q-\omega)^3 (\omega-q_-) (\omega-q_+) D} \,,
\end{split}
\end{equation}
where $D$ is the quantity introduced in eq (\ref{eqdefD}). 

We can compute the entropy using the gauge-invariant formula proposed in \cite{Elgood:2020nls} which we report for completeness 
\begin{equation}
	\label{eq:Waldentropyformula}
	S_W
	=
	\frac{g_{s}^{2}}{8G_{N}^{(10)}}
	\int_{\mathcal{BH}}
	e^{-2\hat\phi}
	\left\{
	\left[
	\hat \star ({\hat e}^{\hat a}\wedge {\hat e}^{\hat b})
	+\frac{\alpha'}{2}\star \hat R^{(0)}_{(-)}{}^{\hat a \hat b}
	\right]{\hat n}_{\hat a\hat b}
	+(-1)^{d}\frac{\alpha'}{2}\Pi_{n}\wedge \hat\star {\hat H}^{(0)}
	\right\}\,.
\end{equation}

\noindent
$\mathcal{BH}$ stands for the horizon bifurcation surface. $\Pi_{n}$ is the vertical Lorentz momentum map associated to the binormal to the Killing horizon, $\hat n^{\hat a\hat b}$, defined by the property

\begin{equation}
	d\Pi_{n}
	\stackrel{\mathcal{BH}}{=}
	{\hat R}^{(0)}_{(-)}{}^{\hat a\hat b}{\hat n}_{\hat a\hat b}\,.
\end{equation}
In the frame that we are using
\begin{equation}
	\Pi_{n}
	\stackrel{\mathcal{BH}}{=}
	{\hat \Omega}^{(0)}_{(-)}{}^{\hat a\hat b}{\hat n}_{\hat a\hat b}\,.
\end{equation}
Replacing our solution we obtain 
\begin{equation}
	S_W = \frac{\pi}{G_N^{(4)}} \sqrt{(q_+ -w)(q_--w)\left[(q-w)^2 + \alpha' \Delta_S\right]} \,,
\end{equation}
with 
\begin{equation}
	\Delta_S = \frac{5 q^2 - 9 q \omega + 3 \omega^2}{10q^2 - 15 q \omega + 5 \omega^2 } + \frac{q \beta_0 \beta_{\mathcal{H}} (10 q^2 - 19 q \omega + 8 \omega^2)}{20(q-\omega)^2(2q-\omega)}+ \frac{q_- q_+ \beta_+ \beta_-}{4 (q_+ - \omega)(q_- - \omega)} \,.
\end{equation}

\subsection{First Law and Smarr Formula}

We want to verify if the thermodynamic quantities we computed satisfy the first law. In order to express the variation of the entropy in term of the physical quantity, we first determine the variation of the mass and charges with respect to the variation of the parameters $q_i$, $w$, $k_\infty$, $\ell_\infty$, $\phi_\infty$ and $\alpha'$, assuming that $ G_N^{(4)} $ is a fixed constant
\begin{subequations}
\begin{align}
	\delta Q_0 = \; & - \frac{g_s^{(4)} \ell_\infty}{4 G_N^{(4)}} \left[\frac{2 q - \omega}{2 q \, \beta_0}\delta q - \frac{1}{2 \beta_0} \delta \omega \right] + Q_0 \, \delta \phi_\infty + Q_0 \, \frac{\delta \ell_\infty }{\ell_\infty}\,, \\
	\delta Q_\mathcal{H} = \; & - \frac{g_s^{(4)} \ell_\infty^{-1}}{4 G_N^{(4)}} \left[\frac{2 q - \omega}{2 q\, \beta_\mathcal{H}}\delta q - \frac{1}{2 \beta_\mathcal{H}} \delta \omega \right]  + Q_\mathcal{H} \, \delta \phi_\infty - Q_\mathcal{H} \, \frac{\delta \ell_\infty }{\ell_\infty} \,, \\
	\delta Q_- = \; & \frac{k_\infty^{-1}}{4 G_N^{(4)} g_s^{(4)}} \left[\frac{2 q_- - \omega}{2 q_- \beta_-}\delta q_- - \frac{1}{2 \beta_-} \delta \omega \right] - Q_- \, \delta \phi_\infty - Q_- \, \frac{\delta k_\infty}{k_\infty} \,, \\
	\delta Q_+ = \; & \frac{k_\infty}{4 G_N^{(4)} g_s^{(4)}} \left[\frac{2 q_+ - \omega}{2 q_+ \beta_i}\delta q_+ - \frac{1}{2 \beta_+} \delta \omega \right] - Q_+ \, \delta \phi_\infty + Q_+ \, \frac{\delta k_\infty}{k_\infty} \,, \\
	\begin{split}
	\delta M = \; &  \frac{1}{4 G_N^{(4)}} \bigg\{\left[2 + \alpha'\frac{2(1-s_0 s_\mathcal{H}) }{5(2q - \omega)^2}\right]\delta q + \delta q_- + \delta q_+  \\
	& + \left[-2 + \alpha'\frac{(-1 + s_0 s_\mathcal{H})}{5(2q - \omega)^2}\right] \delta \omega + \frac{(-1+s_0 s_\mathcal{H})}{10 q - 5 \omega} \delta\alpha' \bigg\} \,.
	\end{split}
\end{align}
\end{subequations}
We then express the variations $\delta q_i $ and $\delta \omega$ in term of the variations of the physical charges and mass. Finally, we replace them in the variation of the entropy expressed in terms of $\delta q_i $, $\delta \omega$ and $\delta \alpha'$. We obtain
\begin{equation}
	\delta S = \frac{1}{T} \, \delta M - \frac{\Phi_i}{T} \,\delta Q_i - \frac{\Phi_{\alpha'}}{T} \, \delta \alpha' - \frac{\mathcal{Q}_k}{T} \delta k_\infty - \frac{\mathcal{Q}_\ell}{T} \delta \ell_\infty - \frac{\mathcal{Q}_\phi}{T} \delta \phi_\infty \,.
\end{equation} 

The temperature $T$ appearing in the first law matches the Hawking entropy (\ref{eqthawking}), providing an highly non-trivial check that the gauge invariant entropy formula proposed by \cite{Elgood:2020nls} is the correct one to use (repeating the same computation with the standard Iyer-Wald prescription we do not recover the Hawking temperature). 

The coefficients $\Phi_\pm$ match the electrostatic potentials of \cite{Elgood:2020nls}, with the subtlety that they must be computed with the fields in the modified Einstein normalization (it will be relevant in order to have  a canonically normalized scalar charge). They are defined by
\begin{equation}
	\Phi_+ 	\stackrel{\mathcal{BH}}{=}  -\iota_t A^z_E \,, \qquad  \Phi_-	\stackrel{\mathcal{BH}}{=}  -\iota_t C^{(1)}_{z \, E} \,.
\end{equation}	
The coefficients $\Phi_{0,\mathcal{H}}$ match the magnetic potentials of \cite{Ortin:2022uxa}. They are defined as the electrostatic potential of the dual gauge fields. Again, they must be computed with the fields in the modified Einstein normalization
\begin{equation}
	\Phi_{\mathcal{H}} 		\stackrel{\mathcal{BH}}{=}  -\iota_t A^{w}_E{}^D \,, \qquad  \Phi_0		\stackrel{\mathcal{BH}}{=}  -\iota_t C^{(1)}{}^D_{\omega \, E} \,.
\end{equation}	
The simplest way to compute them is evaluating first $\Phi_0$ and then obtaining $\Phi_{\mathcal{H}}$ performing a T-duality transformation. In order to compute  $\Phi_0$ we could dualize directly $C^{(1)}{}^D_\omega$. However, it is simpler to perform the dualization in 10d and then perform a dimensional reduction. We start therefore dualizing the KR field directly in in 10d 
\begin{equation}
	\hat{H}^{(7)} = e^{-2 \hat{\phi}} \hat{\star} \, \hat{H} \,.
\end{equation}
We have then\footnote{The dimensional reduction is straightforward because now the Bianchi identity is just $d \mathcal{H}^{(7)} = 0$ and the only non vanishing components are those with the form $\mathcal{H}^{(7)}_{\mu \nu m y^1 \dots y^4}$. In particular, in the relation between higher and lower dimensional fields there are no explicit $\alpha'$ corrections.}
\begin{equation}
	G^{(1)}{}^D_\omega = \frac{1 }{2} \hat{H}^{(7)}_{\mu \nu \bar{z} y^1\dots y^4} \, dx^\mu \wedge dx^\nu \,,
\end{equation}
and the modified Einstein frame field strength 
\begin{equation}
	G^{(1)}_{w \, E}{}^D = G^{(1)}_{w}{}^D \, e^{\phi_\infty} \,.
\end{equation}
The explicit expressions of the potentials are
\begin{subequations}
	\begin{align}
		\Phi_+ & = \frac{k_\infty^{-1}g_s^{(4)}}{\beta_+}\left(1+\alpha' \Delta_{\Phi+}\right) \,, \\
		\Phi_- & =  \frac{k_\infty g_s^{(4)}}{\beta_-}\left(1+\alpha' \Delta_{\Phi-}\right) \,, \\
		\Phi_0 & =  \frac{\ell_\infty^{-1}}{g_s^{(4)} \, \beta_0}\left(1+\alpha' \Delta_{\Phi0}\right) \,, \\
		\Phi_\mathcal{H} & = \frac{\ell_\infty}{g_s^{(4)} \,\beta_\mathcal{H}}\left(1+\alpha' \Delta_{\Phi\mathcal{H}}\right) \,,
	\end{align}	
\end{subequations} 
with
\begin{subequations}
	\begin{align}
		\begin{split}
		\Delta_{\Phi+} = &  \; \frac{  \omega (q-2 \omega) (\omega-2 q_-) (s_0 s_\mathcal{H}+4)}{10 (q-\omega)^3 D} +\frac{\beta_- \beta_+ q_- \omega}{8 (q-\omega)^2 (q_--\omega) (q_+-\omega)} \,,
		\end{split} \\
		\begin{split}
		\Delta_{\Phi-} = & \; \frac{  \omega (q-2 \omega) (\omega-2 q_+) (s_0 s_\mathcal{H}+4)}{10 (q-\omega)^3 D} +\frac{\beta_- \beta_+ q_+ \omega}{8 (q-\omega)^2 (q_--\omega) (q_+-\omega)} \,,
		\end{split} \\
		\begin{split}
		\Delta_{\Phi0} = & \; s_0 s_\mathcal{H} \bigg[4 \omega^2 \left(3 q^2 +qq_-+q q_++3 q_- q_+\right) \\
		&-8 q \omega \left(q^2+2 q q_-+2 qq_++3 q_- q_+\right)+8 q^2 (q q_-+ qq_++2 q_- q_+) \\
		& +\omega^3 (2 q-q_--q_+)-2 \omega^4\bigg] \frac{1}{40 (q-\omega)^3 (2 q-\omega) D}\\
		&  + \bigg[ -8 \omega^2 \left(9 q^2+18 q q_-+18 q q_++4 q_- q_+\right) \\
		& +4 q \omega \left(2 q^2+19 q q_-+19 qq_++16 q_- q_+\right)-8 q^2 (q q_-+qq_++2 q_- q_+) \\
		& +8 \omega^3 (16 q+7 q_-+7q_+)-48 \omega^4 \bigg]  \frac{1}{40 (q-\omega)^3 (2 q-\omega) D}	 \,,
		\end{split} \\
		\begin{split}
		\Delta_{\Phi\mathcal{H}} = &  \; \Delta_{\Phi0}\,,
		\end{split} 
	\end{align}	
\end{subequations} 
where $D$ is the quantity defined in equation (\ref{eqdefD}). 

At zeroth order in $\alpha'$ the scalar charges $\mathcal{Q}_i$ are related to the numerators of the $1/r$ term in the asymptotic expansion of the scalar fields. In particular, given a 2-derivative theory containing Abelian vector fields and scalars coupled to gravity with the scalar kinetic sector
\begin{equation}
	\frac{1}{16 \pi G_N^{(4)}} \int d^4x \sqrt{g_E} \left[ \frac{1}{2} \mathcal{M}_{xy} \, \partial_\mu \psi^x \, \partial^\mu \psi^y \right] \,,
\end{equation} 
we obtain (see \cite{Gibbons:1996af})
\begin{equation}
	\mathcal{Q}_x = -\frac{1}{4} \mathcal{M}_{x y} \,\Sigma^y \,,
\end{equation}
where $\Sigma^x$ are the scalar charges defined by the expansion
\begin{equation}
	\psi^x = \psi^x_\infty + \frac{G_N^{(4)} \Sigma^x}{r} + \mathcal{O}\left(1/r^2\right) \,.
\end{equation}
In the modified Einstein frame the kinetic term of the scalar sector of HST action has the form 
 \begin{equation}
 	\frac{1}{16 \pi G_N^{(4)}} \int d^4x \sqrt{g_E} \bigg[2\, (\partial \phi)^2 + (\partial \log k )^2 + (\partial \log \ell )^2 \bigg] \,,
 \end{equation} 
and we precisely recover
\begin{subequations}
\begin{align}
	& \mathcal{Q}_\phi \, \delta \phi_\infty = - \Sigma_{\phi} \, \delta \phi_\infty + \mathcal{O}(\alpha') \,, \label{eqChargeD}\\
	& \mathcal{Q}_k \,  \delta k_\infty = - \frac{1}{2 }\Sigma_{\log k} \,  \delta \log k_\infty  + \mathcal{O}(\alpha') \label{eqChargeK}\,, \\
	& \mathcal{Q}_\ell \, \delta \ell_\infty = - \frac{1}{2}\Sigma_{\log \ell} \,  \delta \log \ell_\infty + \mathcal{O}(\alpha') \label{eqChargeL} \,.
\end{align}
\end{subequations}
At first order in $\alpha'$ we do not have a prescription yet we can compare with for HST. Developing a formula for the scalar charges applying the technique of \cite{Ballesteros:2023iqb} to HST will be the goal of a future work and is beyond the scope of this one \cite{MatteoWIP}. The explicit forms of the $\mathcal{Q}$s we get are 
\begin{subequations}
\begin{align}
	\mathcal{Q}_\phi & = \Phi_+ \, Q_+  + \Phi_- \, Q_- - \Phi_0 \, Q_0 - \Phi_\mathcal{H} \, Q_\mathcal{H} \,, \\
	\mathcal{Q}_k & = k_\infty^{-1} \left[-\Phi_+ \, Q_+  + \Phi_- \, Q_- \right] \,, \\
	\mathcal{Q}_\ell & = \ell_\infty^{-1} \left[-\Phi_0 \, Q_0  + \Phi_\mathcal{H} \, Q_\mathcal{H}\right] \equiv  0 \,.
\end{align}
\end{subequations}
It is possible to verify that equations (\ref{eqChargeK}) and (\ref{eqChargeL}) are still valid at first order in $\alpha'$, but (\ref{eqChargeD}) is not satisfied anymore.

Finally, the potential $\Phi_{\alpha'}$ is given by
\begin{equation}
	\begin{split}
		\Phi_{\alpha'} = & \; \frac{1}{G_N^{(4)}} \bigg[\frac{4 q^2(-1+s_0 s_\mathcal{H})-5q(-4+s_0 s_\mathcal{H})\omega - 20\omega^2}{160(q-\omega)^3}  \\ 
		& + \frac{5 q_- q_+ (q-\omega) \omega \beta_- \beta_+}{160 (q-\omega)^3(q_- - \omega)(q_+ - \omega)}\bigg]\,,
	\end{split}
\end{equation}
and it is exactly the one which allows the Smarr formula to be satisfied\footnote{The presence of a potential in the Smarr formula is expected for every independent dimensionful parameter. See for instance \cite{Ortin:2021win,Mitsios:2021zrn,Meessen:2022hcg}.}
\begin{equation}
	M = 2 S_W T_H + \Phi_i Q_i + 2 \Phi_{\alpha'} \,\alpha' \,.
\end{equation}

\section{Discussion} \label{secConclusions}

In this paper we have computed for the first time the first-order $\alpha'$ corrections to a family of  static non-extremal, 4-dimensional, 4-charge black holes with 3 independent charges. If we set all the 4 charges equal we obtain new corrections to several embeddings of the Reissner-Nordstr\"om black hole in string theory. In order to compute such corrections a lot of ingredients turned out to be fundamental:
\begin{itemize}
	\item the lemma proven by \cite{Bergshoeff:1989de} which allows us to simply the EOMs; 
	\item the idea of \cite{Cano:2021nzo} of using T-duality to write a sensible ansatz and constrain it. This allows us to reduce the EOMs to a single fourth order differential equation for a single unknown function;
	\item the technique of \cite{Cano:2022tmn} to solve complicate differential equations.
\end{itemize}
The solutions are fully analytical and we checked that they do not present curvature singularities. 

The second result of this work is a detailed study of the thermodynamics of these solutions. Having access to the asymptotic fall-off of the fields we easily evaluated the mass and the gauge charges. We found that the integration constants of the differential equations can be picked in such a way that the charges expressions at first and zeroth order in $\alpha'$ are the same, i.e. they do not get renormalized. However, we checked that the same can not be done for the mass, finding a perfect agreement with the corrections found in \cite{Cano:2021nzo} in the extremal limit. The entropy has been computed with the revisited Wald method of \cite{Elgood:2020svt, Elgood:2020mdx,Elgood:2020nls} and we checked that the thermodynamic relation $\delta S / \delta M = 1/T$ properly reproduces the Hawking temperature. On top of that, we verified that the extremal limit and the Schwarzschild limit reproduce the previous results of the literature.\footnote{See \cite{Cano:2021nzo, Cano:2019ycn} and references therein.} The current work after \cite{Cano:2019ycn, Cano:2022tmn} verifies once more the entropy formula proposed in \cite{Elgood:2020nls} is the correct one to use. 

With the computed entropy we tested further the first law of black hole mechanics. We found that if we ignore the variations of the moduli, the entropy satisfies the first law with the gauge charges and gauge potentials of \cite{Elgood:2020nls,Ortin:2022uxa}. Instead, if we do consider them, the scalar charges we obtain are not the standard ones, not even at zeroth order in $\alpha'$, unless we apply the formulas proposed in the \textit{modified Einstein normalization} for the gauge fields. Such frame has been first introduced in \cite{Gomez-Fayren:2023wxk} in order to have an effective action with no explicit dependencies on the moduli. In the current work we highlight again the necessity of such frame, verifying that it is fundamental to achieve the correct Dirac quantization of the gauge charges. From the first law we have then been able to extract the scalar charges and we found that the ones associated to the Kaluza-Klein scalars match as expected the poles of the field asymptotic expansion. Instead, the dilaton scalar charge receives $\alpha'$ correction and is not simply given by the pole of the dilaton asymptotic expansion. Finally, all the quantities computed do satisfy the Smarr formula provided that we add a gauge potential term for the dimensionful parameter $\alpha'$, in agreement with what found in \cite{Cano:2022tmn} and the expectation of \cite{Meessen:2022hcg}. 

We conclude with a short outlook of some interesting developments. First, the access to a complete analytic solution would allow us to go beyond all the methods currently available to compute $\mathcal{O}(\alpha')$ corrections to the thermodynamic without a complete solution. Indeed, following the method of \cite{Reall:2019sah,Ma:2023qqj}, the solutions may be used to obtain the $\mathcal{O}(\alpha'{}^2)$ corrections. Second, the entropy formula we obtained may be used to test the interpretation proposed in \cite{Horowitz_1996}. The re-parametrization they proposed and the associated interpretation is still missing a complete proof. Now it would be possible to run a new non-trivial test, testing the $\alpha'$ corrections of the limits under control in the microscopic side. Third, it would be interesting to apply the scalar charge definition of \cite{Ballesteros:2023iqb} to HST and deepen our understanding of the origin of the corrections to the standard scalar charge definition in the case of the dilaton. Work in this direction is already in progress \cite{MatteoWIP}. Finally, this solution together with the one of \cite{Cano:2022tmn} may be used to derive explicitly the corrections to the thermodynamics of the configurations considered in \cite{Chen:2021dsw}. Work in this direction is in progress \cite{MatteoWIP2}.

\section*{Acknowledgments}
I would like to thank Tom\'as Ort\'\i{}n, Alejandro Ruip\'erez, Pablo A. Cano and Romina Ballesteros for interesting conversations. This work has been supported by the fellowship LCF/BQ/DI20/11780035 from ``La Caixa'' Foundation (ID
100010434). This work has been supported in part by the MCI, AEI, FEDER (UE) grants PID2021-125700NBC21 (“Gravity, Supergravity and Superstrings” (GRASS)), and IFT Centro de Excelencia Severo Ochoa CEX2020-001007-S.

\appendix

\section{The Heterotic Superstring effective action}
\label{sec-heteroticalpha}

For the sake of self-consistency, we briefly describe the 10-dimensional Heterotic String Theory (HST) effective action to first order in $\alpha'$ and its equations of motion (EOMs). The action is essentially the one of
Ref.~\cite{Bergshoeff:1989de} adapted to the conventions of
\cite{Ortin:2015hya} \footnote{The $\alpha'$ corrections to that action we first studied
	in Refs.~\cite{Gross:1986mw,Metsaev:1987zx,Bergshoeff:1989de}. More recent,
	and relevant, studies of those corrections can be found in
	Refs.~\cite{Chemissany:2007he,Baron:2017dvb,Baron:2018lve}.}.

The HST effective action describes the massless degrees of freedom of the HST\footnote{ In this appendix we will work only with 10 dimensional fields, therefore we suppress the notation with hats. }:
the Zehnbein $e^{a}=e^{a}{}_{\mu}dx^{\mu}$, the Kalb-Ramond (KR)
2-form $B=\tfrac{1}{2}B_{\mu\nu}dx^{\mu}\wedge dx^{\nu}$, the dilaton $\phi$
and the Yang-Mills (YM) field $A^{A}=A^{A}{}_{\mu}dx^{\mu}$ ($A,B,C,\ldots$ take
values in the Lie algebra of the gauge group).

In order to conveniently describe the HST action and its EOMs we introduce some objects. Given the (torsionless, metric-compatible) Levi-Civita spin
connection 1-form $\omega^{a}{}_{b}=\omega_{\mu}{}^{a}{}_{b}dx^{\mu}$ which satisfies in our convention the Cartan structure equations 
\begin{equation}
	\mathcal{D}e^{a}\equiv de^{a}-\omega^{a}{}_{b}\wedge e^{b}=0\,,
\end{equation}
and has curvature 2-form
\begin{equation}
	{R}{}^{{a}}{}_{{b}} = 
	d \omega{}^{{a}}{}_{{b}}
	- {\omega}{}^{{a}}{}_{{c}}
	\wedge  
	{\omega}{}^{{c}}{}_{{b}}\,,
\end{equation}
we define two torsionful spin connections
\begin{equation}
	\Omega_{(\pm)}^{(0)}{}^a{}_b =	{\omega}^{{a}}{}_{{b}}
	\pm
	\tfrac{1}{2}{H}^{(0)}_{{\mu}}{}^{{a}}{}_{{b}}dx^{{\mu}}\,,
\end{equation}
where ${H}^{(0)}$ is the zeroth order field strength of the KR 2-form $B$
\begin{equation}
	{H}^{(0)} = d B \,.
\end{equation}
From the torsionful spin connection $\Omega_{(-)}^{(0)}{}^a{}_b$ we can build the associated curvature 2-form and the Lorentz-Chern-Simons 3-form
\begin{subequations}
		\begin{align}
		{R}^{(0)}_{(-)}{}^{{a}}{}_{{b}}
		& =
		d {\Omega}^{(0)}_{(-)}{}^{{a}}{}_{{b}}
		- {\Omega}^{(0)}_{(-)}{}^{{a}}{}_{{c}}
		\wedge  
		{\Omega}^{(0)}_{(-)}{}^{{c}}{}_{{b}}\,,
		\\[2mm]
		{\omega}^{{\rm L}\, (0)}_{(-)} 
		& =   
		d{\Omega}^{ (0)}_{(-)}{}^{{a}}{}_{{b}} \wedge 
		{\Omega}^{ (0)}_{(-)}{}^{{b}}{}_{{a}} 
		-\tfrac{2}{3}
		{\Omega}^{ (0)}_{(-)}{}^{{a}}{}_{{b}} \wedge 
		{\Omega}^{ (0)}_{(-)}{}^{{b}}{}_{{c}} \wedge
		{\Omega}^{ (0)}_{(-)}{}^{{c}}{}_{{a}}\,.  
	\end{align}
\end{subequations}
Analogously, we can define the curvature 2-form and the Chern-Simons 3-form of the Yang-Mills field 
\begin{subequations}
	\begin{align}
			{F}^{A}	& =	d{A}^{A}+\tfrac{1}{2} f_{BC}{}^{A}{A}^{B}\wedge{A}^{C}\,, \\[2mm]
		{\omega}^{\rm YM}	& = dA_{A}\wedge {A}^{A}+\tfrac{1}{3}f_{ABC}{A}^{A}\wedge{A}^{B}\wedge{A}^{C}\,,
	\end{align}
\end{subequations}
where we have used the Killing metric of the gauge group's
Lie algebra in the relevant representation to lower the indices. We define now the first order field strength of the KR 2-form
\begin{equation}
	\label{eq:H1def}
	H^{(1)}
	= 
	d{B}
	+\frac{\alpha'}{4}\left({\omega}^{\rm YM}+{\omega}^{{\rm L}\, (0)}_{(-)}\right)\,.    
\end{equation}
Notice that the $H^{(1)}$ field strength is invariant under both Yang-Mills and local-Lorentz gauge transformations because they induce a compensating Nicolai-Townsend transformations of $B$. We finally introduce  the so-called ``$T$-tensors'', which can be used to conveniently encode the explicit $\alpha'$ corrections in the action, in the equations of motion and in the Bianchi identity of the Kalb-Ramond 2-form
\begin{equation}
	\label{eq:Ttensors}
	\begin{array}{rcl}
		{T}^{(4)}
		& \equiv &
		\dfrac{\alpha'}{4}\left[
		{F}_{A}\wedge{F}^{A}
		+
		{R}_{(-)}{}^{{a}}{}_{{b}}\wedge {R}_{(-)}{}^{{b}}{}_{{a}}
		\right]\,,
		\\
		& & \\ 
		{T}^{(2)}{}_{{\mu}{\nu}}
		& \equiv &
		\dfrac{\alpha'}{4}\left[
		{F}_{A}{}_{{\mu}{\rho}}{F}^{A}{}_{{\nu}}{}^{{\rho}} 
		+
		{R}_{(-)\, {\mu}{\rho}}{}^{{a}}{}_{{b}}{R}_{(-)\, {\nu}}{}^{{\rho}\,  {b}}{}_{{a}}
		\right]\,,
		\\
		& & \\    
		{T}^{(0)}
		& \equiv &
		{T}^{(2)\,\mu}{}_{{\mu}}\,.
		\\
	\end{array}
\end{equation}

The  string-frame HST effective action is then, to first order in $\alpha'$,
\begin{equation}
	\label{heterotic}
	S
	=
	\frac{g_{s}^{2}}{16\pi G_{N}^{(10)}}
	\int d^{10}x\sqrt{|{g}|}\, 
	e^{-2{\phi}}\, 
	\left\{
	{R} 
	-4(\partial{\phi})^{2}
	+\tfrac{1}{12}{H}^{(1)\, 2}
	-\tfrac{1}{2}T^{(0)}
	\right\}\,,
\end{equation}

\noindent
where $R$ is the Ricci scalar of the string-frame metric
$g_{\mu\nu}=\eta_{ab}e^{a}{}_{\mu}e^{b}{}_{\nu}$, $G_{N}^{(10)}$ is the
10-dimensional Newton constant, $g_{s}$ is the HST coupling constant (the
vacuum expectation value of the dilaton $e^{<\phi>}$ which we will identify
with the asymptotic value of the dilaton $e^{\phi_{\infty}}$ in
asymptotically-flat black-hole solutions). The 10-dimensional Newton constant, the string length $\ell_s = \sqrt{\alpha'}$ and the string coupling
constant $g_s$ are related by
\begin{equation}
	\label{eq:10dNewtonconstant}
	G_{N}^{(10)} = 8\pi^{6}g_{s}^{2}\ell_{s}^{8}\,.  
\end{equation}
Notice that the expression of the action we are considering contains terms of order $\mathcal{O}(\alpha'{}^2)$ that we must drop in the computation of the EOMs. However, such expression of the action has the advantage of being manifestly invariant under gauge transformations. Dropping the $\mathcal{O}(\alpha'{}^2)$ terms it would be invariant up to terms of order $\mathcal{O}(\alpha'{}^2)$.

Now we want to compute the EOMs. The naive variation of the action (\ref{heterotic}) leads to very complicated equations of motion which contain terms with higher derivatives. However, it can be shown that all of them come from the variation of the torsionful spin connection and they are proportional to the zeroth order EOMs (see the lemma proven in Ref.~\cite{Bergshoeff:1989de}). 
Therefore, in order to compute the corrections to a solution of the zeroth order EOMs, we can consistently ignore those terms. The variation of the HST action with $\delta \Omega_{(-)}{}^a{}_b = 0$ leads then to a set of EOMs that can be written in the form
\begin{subequations}
\begin{align}
	R_{\mu\nu} -2\nabla_{\mu}\partial_{\nu}\phi
	+\tfrac{1}{4}{H}_{\mu}{}^{\rho\sigma}{H}_{\nu\rho\sigma}
	-T^{(2)\,}_{\mu\nu} = 0 	\label{eq:eq1} \,, \\[2mm]
	(\partial \phi)^{2} -\tfrac{1}{2}\nabla^{2}\phi
	-\tfrac{1}{4\cdot 3!}{H}^{2}
	+\tfrac{1}{8}T^{(0)} = 0 	\label{eq:eq2}  \,, \\[2mm]
	\nabla_{\mu}\left(e^{-2\phi}H^{\mu\nu\rho}\right) = 0 	\label{eq:eq3} \,, \\[2mm]
	\alpha' e^{2\phi}\nabla_{(+)\, \mu}\left(e^{-2\phi}F^{A\, \mu\nu}\right) = 
	0 	\label{eq:eq4} \,,
\end{align}
\end{subequations}
where $\nabla_{(+)\, \mu}$ is the covariant derivative which is covariant with respect to $\Omega_{(+)}^{(0)}{}^a{}_b$ and the YM gauge transformation. Notice that the YM field can be consistently truncated. Finally, the KR field strength satisfies the Bianchi identity
\begin{equation}
	d H^{(1)} - \frac{1}{3}T^{(4)} = 0 \,.
\end{equation}

\section{Relation between $10d$ and $4d$ fields at first order in $\alpha'$}\label{sec10to4d}
We start recalling the relations between the 10-dimensional fields (indicated with an hat) and the $(10-d)$-dimensional fields obtained via a $\text{T}^n$ dimensional reduction at first order in $\alpha'$. They are essentially those of \cite{Ortin:2020xdm}. Notice that we consistently truncated the YM fields. We decompose the indexes as $\hat{\mu} = (\mu, m)$.
\begin{subequations}
\begin{align}
	& {g}_{\mu\nu} = \hat{g}_{\mu\nu}-\hat{g}^{mn}\hat{g}_{\mu m}\hat{g}_{\nu n} \,,\\[2mm]
	& A^n{}_\mu = \hat{g}^{nm} \hat{g}_{m\mu}  \,, \\[2mm]
	& G_{mn} = - \hat{g}_{mn}  \,, \\[0mm]
	& { \phi} = \hat{\phi} - \frac{1}{2}\log \det ( \hat{g}_{mn} ) \,, \\
	& {B}^{(1)}{}_{\mu\nu} = \hat{B}_{\mu\nu} + \hat{g}^{mn} \hat{g}_{m[\mu}\hat{B}_{\nu]n} - \frac{\alpha'}{4}\left(\hat{\Omega}_{(-)}^{(0)}{}_m{}^{\hat{a}}{}_{\hat{b}} \hat{\Omega}_{(-)}^{(0)}{}_{[\mu|}{}^{\hat{b}}{}_{\hat{a}}\right) \hat{g}^{mn} \hat{g}_{|\nu]n} \,, \\
	\begin{split}
	& {C}^{(1)}{}_{m \mu} = \hat{B}_{\mu m} + \left[\hat{B}_{mn}- \frac{\alpha'}{4}\left(\hat{\Omega}_{(-)}^{(0)}{}_{m}{}^{\hat{a}}{}_{\hat{b}} \hat{\Omega}_{(-)}^{(0)}{}_{n}{}^{\hat{b}}{}_{\hat{a}}\right)\right] \hat{g}^{np}\hat{g}_{p \mu}  \\[2mm]
	& \qquad \qquad - \frac{\alpha'}{4} \left(\hat{\Omega}_{(-)}^{(0)}{}_m{}^{\hat{a}}{}_{\hat{b}} \hat{\Omega}_{(-)}^{(0)}{}_{\mu}{}^{\hat{b}}{}_{\hat{a}}\right) \,,
	\end{split} \\[2mm]
	& {K}^{(1)}{}_{m n} = \hat{B}_{mn} - \frac{\alpha'}{4}\left(\hat{\Omega}_{(-)}^{(0)}{}_{[m|}{}^{\hat{a}}{}_{\hat{b}} \hat{\Omega}_{(-)}^{(0)}{}_{|n]}{}^{\hat{b}}{}_{\hat{a}}\right)  \,. 
\end{align} 
\end{subequations}
where $G_{mn}$ and $ {K}^{(1)}{}_{m n}$ are matrices of scalars, $A^n{} $ and $C^{(1)}_{m}$ are gauge vectors. $g_{\mu\nu}$, $B^{(1)}$ and $\phi$ are the lower dimensional metric, KR 2-form and dilaton. $g^{mn}$ is the inverse of $g_{mn}$.

In the setup considered in this work we have a $\text{T}^6$ compactification with a trivial torus $\text{T}^4$ and two independent $S^1$ parametrized by $w$ and $z$. Omitting the $\text{T}^4$ indexes, $G_{mn}$ and $K^{(1)}_{mn}$ have the explicit form
\begin{equation}
	G_{mn} = \begin{pmatrix}
		G_{ww} &  0 \\ 0 & G_{zz}  
	\end{pmatrix} \equiv  \begin{pmatrix}
\ell^2 &  0 \\ 0 & k^2
\end{pmatrix} \,, \qquad K^{(1)}_{mn} = 0 \,.
\end{equation}

In order to study T-duality transformations along the $w$ and $z$ direction, following \cite{Elgood:2020xwu} it is useful to introduce the combinations 
\begin{subequations}\label{eqdefl1k1}
\begin{align}
	& \ell^{(1)} = \ell\left[1 - \frac{\alpha'}{4}\left(\hat{\Omega}_{(-)}^{(0)}{}_{w}{}^{\hat{a}}{}_{\hat{b}} \hat{\Omega}_{(-)}^{(0)}{}_{w}{}^{\hat{b}}{}_{\hat{a}}\right)\hat{g}^{ww}\right] \,,  \\
	& k^{(1)} = k\left[1 - \frac{\alpha'}{4}\left(\hat{\Omega}_{(-)}^{(0)}{}_z{}^{\hat{a}}{}_{\hat{b}} \hat{\Omega}_{(-)}^{(0)}{}_{z}{}^{\hat{b}}{}_{\hat{a}}\right)\hat{g}^{zz}\right] \,.
\end{align}
\end{subequations}

\section{T-duality constraints} \label{secTdual}
We follow the same logic of \cite{Cano:2021nzo}. Despite the fact that we are dimensionally reducing 10d HST on a 6-dimensional torus, the compactification is trivial in 4 directions and the remaining ones, parameterized by $z$ and $w$, are simply 2 independent 1-dimensional reductions. In order to implement T-duality transformations along $w,z$ we can profit then of the results of \cite{Elgood:2020xwu}. Performing a T-duality transformation, some of the lower dimensional fields must
be invariant
\begin{equation}\label{eqtdual1}
T_{z,w}:\qquad	ds^2_{E} \leftrightarrow ds^2_{E} \,, \qquad e^{-2\phi}\leftrightarrow e^{-2\phi} \,,
\end{equation}
and the remaining ones must transform as follow
\begin{subequations}\label{eqtdual2}
\begin{align}
&T_z: \qquad 	C^{(1)}_z \leftrightarrow A^z \,, \qquad k \leftrightarrow 1/k^{(1)} \,,\\
&T_w: \qquad 	C^{(1)}_w \leftrightarrow A^w \,, \qquad \ell \leftrightarrow 1/\ell^{(1)}  \,.
\end{align}
\end{subequations}
At zeroth order it is simple to verify that these transformations can be implemented simply acting on the constants $q_i$, $k_\infty$ and $\ell_\infty$. We have indeed 
\begin{subequations} \label{eqmapTdual}
\begin{align}
	&T_z: \qquad 	q_+ \leftrightarrow q_- \,, \qquad \beta_+ \leftrightarrow \beta_- \,, \qquad k_\infty \leftrightarrow 1/k_\infty \,,\\
	&T_w: \qquad 	q\leftrightarrow q \,,  \qquad \beta_{\mathcal{H}} \leftrightarrow \beta_0 \,, \qquad \ell_\infty \leftrightarrow 1/\ell_\infty  \,.
\end{align}
\end{subequations}
It has been verified in several solutions that whenever the $\alpha'$ corrections do not modify the asymptotic values of the fields and the form of the poles of the functions relevant for the definitions of the physical charges (in this case, $\delta \mathcal{Z}_{h0}$, $\delta\mathcal{Z}_{h-}$ and $\delta\mathcal{Z}_+$), the zeroth order map between parameters is valid at first order too. In order to obtain non-trivial constraints we assume that this is true in this case too.\footnote{Notice that this assumption can not generate any loophole in the main result of the paper, which is building the BH solutions. The constraints we derive in this section are used to find solutions of the EOMs. If the solutions turn out not to fulfill the assumption they are still valid BH solutions. Anyway, one can verify that all the solutions we get fulfill this property.}

We impose now that our ansatz (\ref{ansatz4d}) satisfies the transformation rules (\ref{eqtdual1}) and (\ref{eqtdual2}). Imposing the invariance of the Einstein metric and the dilaton we obtain the invariance of the string frame metric. From the invariance of the string frame metric we obtain
\begin{equation}\label{eqtdualds}
	T_m : \quad \frac{W_{tt}}{\mathcal{Z}_+\mathcal{Z}_-} \leftrightarrow  \frac{W_{tt}}{\mathcal{Z}_+\mathcal{Z}_-} \,, \quad  \mathcal{Z}_0\mathcal{Z}_{\mathcal{H}} \leftrightarrow  \mathcal{Z}_0\mathcal{Z}_{\mathcal{H}} \,, \quad W_{rr} \leftrightarrow W_{rr}\,.
\end{equation}
Combing these relations with the invariance of the dilaton we further get
\begin{equation}\label{eqtdualdila}
	T_m: \quad \frac{c_\phi}{r^2 \mathcal{Z}_{h-}'} W_{tt} \left(\frac{\mathcal{Z}_{h-}}{\mathcal{Z}_{-}}\right)^2\leftrightarrow   \frac{c_\phi}{r^2 \mathcal{Z}_{h-}'} W_{tt} \left(\frac{\mathcal{Z}_{h-}}{\mathcal{Z}_{-}}\right)^2 \,.
\end{equation}
From the transformation property of $A^z$  we obtain 
\begin{equation} \label{eqtdualzp}
	T_z[\mathcal{Z}_+] = \mathcal{Z}_{h-}\left(1-\alpha' \frac{\Delta_C}{\beta_-}\right)\,,
\end{equation}
where we used the transformation properties of $\beta_\pm$ of equation (\ref{eqmapTdual}). Using the invariance of $\Delta_C$ under $T_z$ we can invert the relation and obtain 
\begin{equation}\label{eqtdualzmh}
	T_z[\mathcal{Z}_{h-}] = \mathcal{Z}_{+}\left(1+\alpha' \frac{\Delta_C}{\beta_-}\right)\,.
\end{equation}
Combining the transformation property of $k$ with the relation (\ref{eqtdualzp}) we get
\begin{equation}\label{eqtdualzm}
	T_z [\mathcal{Z}_-] = \mathcal{Z}_+ \frac{\mathcal{Z}_{h-}}{\mathcal{Z}_-}\left(1 +2\alpha'\Delta_k-\alpha'\frac{\Delta_C}{\beta_-}\right)\,.
\end{equation}
Combining equations (\ref{eqtdualds}), (\ref{eqtdualzp}) and (\ref{eqtdualzm}) we obtain the transformation properties of $W_{tt}$
\begin{equation}\label{eqtdualwtt}
	T_z[W_{tt}] = W_{tt}\left(\frac{\mathcal{Z}_{h-}}{\mathcal{Z}_-}\right)^2\left(1+2\alpha' \Delta_C - 2\alpha' \frac{\Delta_C}{\beta_-}\right)\,.
\end{equation}
Replacing (\ref{eqtdualzmh}), (\ref{eqtdualzm}) and (\ref{eqtdualwtt}) into (\ref{eqtdualdila}) we can obtain a new expression for $\mathcal{Z}_-$ 
\begin{equation} \label{eqtdualzmbis}
	\mathcal{Z}_- = \mathcal{Z}_{h-}\left(1-\alpha' \frac{\Delta_C}{\beta_+}+\alpha' \Delta_k\right) \sqrt{\frac{c_\phi T_z[\mathcal{Z}_{h-}']}{T_z[c_\phi]\mathcal{Z}_{h-}'}} \,.
\end{equation}
The transformation property of $A^w$ is trivial and does not produce any constraint. Finally, the transformation property of $\ell$ combined with the invariance of the metric produces
\begin{subequations}
\begin{align}
	& T_w [\mathcal{Z}_0] = \mathcal{Z}_{\mathcal{H}} (1-\alpha' \Delta_\ell) \,, \\[2mm]
	& T_w [\mathcal{Z}_{\mathcal{H}}] = \mathcal{Z}_0 (1 + \alpha' \Delta_\ell) \,.
\end{align}
\end{subequations}

\section{Charges and Mass Identification}\label{secChargesandMass}

Within this section we assume the ansatz for the $\delta \mathcal{Z}$s and the $\delta W$s
 \begin{equation}
	\delta \mathcal{Z}_i = \sum_{k>0} \frac{d_i^{(k)}}{r^k} \,, \qquad \delta \mathcal{Z}_{hi} = \sum_{k>0} \frac{d_{h-}^{(k)}}{r^k} \,, \qquad
	\delta W_{j} = \sum_{k>0} \frac{d_{wj}^{(k)}}{r^k}\,.
\end{equation}
From the expansion of the 4d dilaton we obtain 
\begin{equation}
	e^{-2\phi} \sim \frac{c_\phi}{q_-+\alpha' d_{h-}^{(1)}} + \mathcal{O}(1/r) \,,
\end{equation}
which gives
\begin{equation}
	c_\phi = e^{-2\phi_\infty} \left(q_- + \alpha' d_{h-}^{(1)}\right)\,.
\end{equation}
Using the definitions for the electric charges
\begin{subequations}
\begin{align}
	Q_+ & = \frac{1}{16 \pi G_N^{(4)}} \int_{S^2_\infty} e^{-2\phi}k^2_E\star_{E} F^z_E \,, \\
	Q_- & = \frac{1}{16 \pi G_N^{(4)}} \int_{S^2_\infty} e^{-2\phi}k^{-2}\star_E G_{z\, E}^{(1)}  \,, 
\end{align}
\end{subequations}
and the field definitions (\ref{ansatz4d}) we obtain
\begin{subequations}
	\begin{align}
		Q_+ & = \frac{k_\infty}{4 G_N^{(4)}g_s^{(4)}} \beta_{+}\left(q_++\alpha' d_{+}^{(1)} \right)  \,, \\
		Q_- & = \frac{k_\infty^{-1}}{4 G_N^{(4)}g_s^{(4)}} \beta_{-} \left(q_-+\alpha' d_{h-}^{(1)} \right) \,, \\
	\end{align}
\end{subequations}
For the magnetic charges we have
\begin{subequations}
	\begin{align}
		Q_\mathcal{H} & = -\frac{1}{16 \pi G_N^{(4)}} \int_{S^2_\infty}  F^w_E \,, \\
		Q_0 & = -\frac{1}{16 \pi G_N^{(4)}} \int_{S^2_\infty}  G_{w\,E}^{(1)} \,. 
	\end{align}
\end{subequations}
The charge $Q_\mathcal{H}$ can be computed using the ansatz (\ref{ansatz4d}) 
\begin{equation}
	Q_\mathcal{H} = \frac{\ell_\infty^{-1} g_s^{(4)}}{4 G_N^{(4)}} \, \beta_\mathcal{H}\,q \,. 
\end{equation}
The charge $Q_0$ instead cannot be computed from the ansatz (\ref{ansatz4d}). The form of $C_w^{(1)}$ is indeed obtained imposing that the asymptotic charge of $C_w^{(1)}$ does not receive corrections. In order to verify such condition we have to utilize the expression which is relating $G^{(1)}_w$ and the higher dimensional fields. Given that we are dimensionally reducing our fields with two independent $\text{S}^1$ compactifications we can profit of the results obtained in \cite{Elgood:2020xwu}. In particular, the relation 
\begin{equation}
	\tfrac{1}{2} \hat{H}_{\mu \nu \underline{w}} \, {dx^\mu \wedge dx^\nu} = G_{w}^{(1)}  +\frac{\alpha'}{4}\Delta+\left(\ell^{(1)}-\ell\right)F^{w}\,,
\end{equation}
holds with $\Delta = \tfrac{1}{2}\Delta_{\mu\nu}\,{dx^\mu \wedge dx^\nu}$ a 2-form defined by
 \begin{subequations}
 \begin{align}
 	\Delta_{\mu\nu} & = R_{(-)}{}_{\mu\nu}{}^{\rho\sigma}K^{(+)}{}_{\rho\sigma}-\tfrac{1}{2}K^{(-)}{}_{\mu\rho}K^{(-)}{}_{\nu\sigma}K^{(+)}{}^{\rho\sigma} -4\nabla_{(-)}{}_{[\mu}K^{(-)}{}_{\nu] \rho}\partial^{\rho}\log{k}\,,\\
 	K^{(\pm)}{}_{\mu\nu} & = k^{(1)}F_{\mu\nu}\pm k^{-1}G^{(1)}{}_{\mu\nu}\,.
 \end{align}
 \end{subequations}
We obtain
\begin{equation}
	Q_0  = -\frac{1}{16 \pi G_N^{(4)}} \int_{S^2_\infty}  g_s^{(4)} \hat{H}_{\underline{w}} + \mathcal{O}(1/r^3) = \frac{\ell_\infty g_s^{(4)}}{4 G_N^{(4)}} \, \beta_0\,\left(q + d_{h0}^{(1)}\right) \,.
\end{equation}
Imposing that the charges do not receive $\alpha'$ corrections, we obtain
\begin{equation} \label{eqnorencharge}
	d_{h0}^{(1)} = d_{h-}^{(1)} = d_+^{(1)} = 0\,.
\end{equation}

We focus now on the mass. We define it via the expansion of the $g_{tt,E}$ component of the 4d metric in the modified Einstein frame
\begin{equation}
	g_{tt,E} \sim 1 - \frac{2 G_N^{(4)} M}{r} + \mathcal{O}(1/r^2)
\end{equation}
Using the ansatz (\ref{ansatz4d}) and the conditions (\ref{eqnorencharge}) we obtain
\begin{equation}
\begin{split}
	M = \; & \frac{1}{4 G_N^{(4)}} \bigg\{2 q + q_- + q_+ - 2 \omega \\
	& \hspace{1.5cm} + \alpha'\left[d_0^{(1)}+d_\mathcal{H}^{(1)}+\left(1+\frac{q_--q_+}{2q_--\omega}\right)d_{wt}^{(1)} + d_{wr}^{(1)}\right]\bigg\}\,.
\end{split}
\end{equation}
One would be tempted to fix $d_\mathcal{H}^{(1)}$ in such a way that the mass does not receive $\alpha'$ corrections as well. However, combining this requirement with the constraints imposed by the regularity on $d_{0}^{(1)}$, $d_{wt}^{(1)}$ and  $d_{wr}^{(1)}$ we would obtain that $d_\mathcal{H}^{(1)}$ is divergent in the extremal limit $\omega \rightarrow 0$. Moreover, we notice that fixing $d_{0}^{(1)}$, $d_{wt}^{(1)}$, $d_{wr}^{(1)}$ and assuming that  $d_\mathcal{H}^{(1)}$ is finite in the extremal limit, we recover the same result of \cite{Cano:2021nzo} independently off $d_\mathcal{H}^{(1)}$ 
\begin{equation}
	M_{\text{ext}} = \frac{1}{4 G_N^{(4)}} \left[2 q + q_- + q_+ -  (1-s_0 s_\mathcal{H})\frac{\alpha'}{10q} \right]\,.
\end{equation}
In absence of a precise prescription to identify the mass, we fix $d_{\mathcal{H}}^{(1)}$ in such a way that the mass is given by
\begin{equation}
		M = \frac{1}{4 G_N^{(4)}} \left[2 q + q_- + q_+ -2\omega -  (1-s_0 s_\mathcal{H})\frac{\alpha'}{5 (2q-\omega)} \right]\,.
\end{equation}
This choice is convenient for several reason. First, the integration constants would be well defined in several interesting limits (extremal limit, Reissner-Nordström limit, purely electrically charged limit). Second, the mass has the correct extremal limit already found in \cite{Cano:2021nzo}. Finally, it is worth to notice the $\alpha'$ corrections are invariant in form under the mapping
\begin{equation}
	\tilde{r} = r + \omega\,, \qquad \tilde{\omega} = - \omega \,, \qquad \tilde{q}_i = q_i - \omega \,.
\end{equation} 
The formula of the mass is then the same one would obtain assuming $\omega > 0$ and running the regularization procedure around $r_H = 0$.

\section{Relevant Limits} \label{secLimits}

In this section we describe some relevant limits of our non extremal, 4-charge, solutions and their thermodynamics.  

\subsection{Extremal Limit}

In the limit $\omega = 0$ we obtain the solutions found in \cite{Cano:2021nzo} 
\begin{subequations}
	\begin{align}
		\begin{split}
			& \delta \mathcal{Z}_{h0} = (1+s_0 s_{\mathcal{H}}) \frac{q^2}{4 r (q+r)^3}\,,
		\end{split} \\[2mm]
		\begin{split}
			& \delta \mathcal{Z}_{0} = (-1+s_0 s_{\mathcal{H}})\frac{34q^2 + 23 qr + 6 r^2}{120 q (q+r)^3} + (1+s_0 s_{\mathcal{H}}) \frac{q^2}{4 r (q+r)^3} \,,
		\end{split} \\[2mm]
		\begin{split}
			& \delta \mathcal{Z}_{\mathcal{H}} =  (-1+s_0 s_{\mathcal{H}})\frac{34q^2 + 23 qr + 6 r^2}{120 q (q+r)^3} \,,
		\end{split} \\[2mm]
		\begin{split}
			& \delta \mathcal{Z}_+ = -(1 + s_+ s_-) \frac{q_+ q_-}{4 r(q+r)^2(q_-+r)} \,,
		\end{split} \\[2mm]
		\begin{split}
			& \delta W_{rr} =    (-1+s_0 s_{\mathcal{H}}) \frac{r(4q+r)}{60(q+r)^4}\,,
		\end{split}  \\[2mm]
		\begin{split}
			& 	\delta \mathcal{Z}_{h-} = \delta \mathcal{Z}_- = \delta W_{tt} = 0 \,.
		\end{split}
	\end{align}
\end{subequations}
The thermodynamic quantities take the form
\begin{subequations}
	\begin{align}
		S_W = & \; \frac{\pi}{2G_N^{(4)}} \sqrt{q_- q_+[4 q^2 + \alpha' (2 + s_- s_+ + s_0 s_\mathcal{H})]} \,, \\
		M = & \; \frac{1}{4 G_N^{(4)}} \left[2q + q_- + q_+ + \alpha'\frac{(-1 + s_0 s_\mathcal{H})}{10 q}\right]  \,, \\
		Q_+ = & \; \frac{k_\infty s_+ q_+}{4 G_N^{(4)}g_s^{(4)}} \,, \\
		Q_- = & \; \frac{k_\infty^{-1} s_- q_-}{4 G_N^{(4)}g_s^{(4)}} \,, \\
		Q_0 = & \;  \frac{g_s^{(4)} \ell_\infty \, s_0\,  q}{4 G_N^{(4)}} \,, \\
		Q_\mathcal{H} = & \;  \frac{g_s^{(4)} \ell_\infty^{-1} \, s_\mathcal{H}\,  q}{4 G_N^{(4)}} \,, \\
		T_H = &  \; 0 \,, \\
		\Phi_+ = & \; k_\infty^{-1} s_+ g_s^{(4)} \,,\\	 
		\Phi_- = & \; k_\infty s_-  g_s^{(4)} \,, \\	
		\Phi_0 = & \; \frac{\ell_\infty^{-1}}{g_s^{(4)}} s_0 \bigg[1+\alpha'\left(\frac{1-s_0s_\mathcal{H}}{20 q^2}\right)\bigg] \,, \\
		\Phi_\mathcal{H} = & \;  \frac{\ell_\infty}{g_s^{(4)}} s_\mathcal{H} \bigg[1+\alpha'\left(\frac{1-s_0s_\mathcal{H}}{20 q^2}\right)\bigg] \,, \\	
		\Phi_{\alpha'} = & \; \frac{1}{G_N^{(4)}}\frac{(-1 + s_0 s_\mathcal{H})}{40 q}	\,.
	\end{align}
\end{subequations}
Notice that in term of the quantum numbers $N_i = Q_i \ell_s g_s^{(4)}$ the macroscopic entropy has the standard form found in \cite{Cano:2021nzo} and references therein
\begin{equation}
	S_W = 2 \pi \sqrt{|N_+N_-|\big(|N_0 N_\mathcal{H}|+ 2 + s_- s_+ + s_0 s_\mathcal{H} \big)  } \,.
\end{equation}
Moreover, it matches the microscopic entropy computed in \cite{Sen:2007qy, Kutasov:1998zh, Kraus:2005vz} which is considered to be exact in $\alpha'$.  

\subsection{Reissner-Nordström Limit}
In the limit $q_+ = q_- = q $ most of the $\delta\mathcal{Z}$s and $\delta W$s are not modified except for the implicit dependence on $q_+$ and $q_-$ via $d_{\mathcal{H}}^{(1)}$. We obtain (we write only the functions with an explicit modification)
\begin{subequations}
	\begin{align}
		\delta \mathcal{Z}_{h-} = & -\frac{q}{r^2}d_{\mathcal{H}}^{(1)}-\frac{q^2 s_0 s_{\mathcal{H}} \left(2 q^2-6 q \omega+5 \omega^2\right)}{20 r^2 (q-\omega)^3 (\omega-2 q)}-\frac{q \left(q^3-3 q^2 \omega+5 q \omega^2-5 \omega^3\right)}{10 r^2 (q-\omega)^3 (2 q-\omega)} \,, \\[4mm]
		\begin{split}
			\delta \mathcal{Z}_- = & \; \delta \mathcal{Z}_{h-} + \mathcal{Z}_- \left[\Delta_k - \frac{\Delta_C}{\beta_+}\right] \,,
		\end{split} \\[4mm]
		\begin{split}
			\delta \mathcal{Z}_+ = & \; \delta \mathcal{Z}_{h-} - \mathcal{Z}_+ \frac{\Delta_C}{\beta_+} \,,
		\end{split} \\[4mm]
		\begin{split}
			\delta W_{tt} = & \; -\frac{q s_0 s_{\mathcal{H}} \omega \left(2 q^2-6 q \omega+5 \omega^2\right) (r+\omega)}{20 r^2 (q-\omega)^4 (2 q-\omega)}-\frac{ s_- s_+ q \omega (r+\omega)}{4 r^2 (q+r)^3}  \\
			&  +\frac{\omega}{80 r^2 (q+r)^3 (q-\omega)^4 (2 q-\omega)} \bigg[ -50 r \omega^3 (q+r)^3 \\
			& -10 q^2 \omega (q+r)^3 (2 q+2 r+\omega)-25 \omega^4 (q+r)^3-50 q \omega^3 (q+r)^3 \\
			& + (4 q (q+r)^3 \left(2 q^2-6 q \omega+5 \omega^2\right) (r+\omega) +15 \omega^3 (q+r)^3 (2 q-\omega) \\
			& +10 q \omega^2 (q+r)^3 (4 q+5 r+3 \omega)-20 q r (2 q-\omega) (q-\omega)^4 \\
			& +10 r \omega (2 q-\omega) (q-\omega)^4-10 q \omega (2 q-\omega) (q-\omega)^4\bigg] +\frac{\omega (r+\omega)}{r^2 (q-\omega)}d_{\mathcal{H}}^{(1)} \,,
		\end{split}
	\end{align}
\end{subequations}
with
\begin{equation}
	d_{\mathcal{H}}^{(1)} = \frac{4 q^3 (s_0 s_{\mathcal{H}}-1)-14 q^2 \omega (s_0 s_{\mathcal{H}}-1)+q \omega^2 (16 s_0 s_{\mathcal{H}}-11)+\omega^3 (9-4 s_0 s_{\mathcal{H}})}{40 (q-\omega)^3 (2 q-\omega)} \,.
\end{equation}
The thermodynamic quantities take the form
\begin{subequations}
	\begin{align}
		S_W = & \;  \frac{\pi}{2 G_N^{(4)}}\sqrt{(q-\omega)^2\left[4(q-\omega)^2+\alpha'\Delta_{S}\right]}\,, \\
		\Delta_S = & \; 4 + \frac{(s_+ s_- - 1) q}{q-\omega} + \frac{(s_0 s_\mathcal{H}-1)(10 q^2 - 19 q\omega + 8 \omega^2)}{5(2q-\omega)(q-\omega)} \,, \\
		M = & \; \frac{1}{4 G_N^{(4)}}\left[4 q - 2\omega + \alpha'\frac{(-1 + s_0 s_\mathcal{H})}{5(2q-\omega)}\right] \,, \\
		Q_+ = & \; \frac{k_\infty \, s_+ }{4 G_N^{(4)} g_s^{(4)}} \sqrt{q(q-\omega)} \,, \\
		Q_- = & \; \frac{k_\infty^{-1} \, s_-}{4 G_N^{(4)}g_s^{(4)}}\sqrt{q(q-\omega)} \,, \\
		Q_0 = & \; \frac{g_s^{(4)} \ell_\infty \, s_0}{4 G_N^{(4)}}\sqrt{q(q-\omega)} \,, \\
		Q_\mathcal{H} = & \; \frac{g_s^{(4)} \, \ell_\infty^{-1} \, s_\mathcal{H}}{4 G_N^{(4)}}\sqrt{q(q-\omega)} \,, \\
		T_H = &  \; \frac{-\omega}{4\pi(q-\omega)^2} \left[1+ \alpha' \frac{q (s_- s_+ -1)}{8(q-\omega)^3} - \alpha'\frac{(s_0 s_\mathcal{H}-1)(6q^2 - 9 q\omega + 4\omega^2)}{40 (q-\omega)^3(2q-\omega)}\right] \,, \\
		\Phi_+ = & \; k_\infty^{-1} s_+ g_s^{(4)} \sqrt{\frac{q}{q-\omega}} \bigg[1+\alpha'\left(\frac{5\omega(2q-\omega) s_- s_+ + 2\omega (4+s_0 s_\mathcal{H})(q-2\omega)}{40(q-\omega)^3(2q-\omega)}\right)\bigg]\,,\\	 
		\Phi_- = & \; {k_\infty s_-} g_s^{(4)} \sqrt{\frac{q}{q-\omega}} \bigg[1+\alpha'\left(\frac{5\omega(2q-\omega) s_- s_+ + 2\omega (4+s_0 s_\mathcal{H})(q-2\omega)}{40(q-\omega)^3(2q-\omega)}\right)\bigg] \,, \\	
		\Phi_0 = & \;  \frac{\ell_\infty^{-1}}{g_s^{(4)}} s_0 \sqrt{\frac{q}{q-\omega}} \bigg[1+\alpha'\Delta_{\Phi0}\bigg] \,, \\
		\Delta_{\Phi0} = & \; \frac{q^2(4-4s_0 s_\mathcal{H}) + 4q\omega(-6+s_0s_\mathcal{H})+ \omega^2(24 + s_0 s_\mathcal{H})}{40 (q-\omega)^3(2q-\omega)} \,, \\
		\Phi_\mathcal{H} = & \;  \frac{\ell_\infty}{g_s^{(4)}} s_\mathcal{H} \sqrt{\frac{q}{q-\omega}} \bigg[1+\alpha' 	\Delta_{\Phi0}\bigg] \,, \\	
		\Phi_{\alpha'} = & \; \frac{1}{G_N^{(4)}}\bigg\{\frac{4q^2(-1+s_0 s_\mathcal{H})-5q(-4+s_0s_\mathcal{H})\omega - 20 \omega^2)}{160 (q-\omega)^3}+ \frac{\omega q s_+ s_-}{32(q-\omega)^3 }\bigg\}\,.
	\end{align}
\end{subequations}
We want to compare the entropy with previous results of the literature for non-extremal black holes. In \cite{Cano:2019ycn} have been computed the $\alpha'$ corrections for a Reissner-Nordstr\"om (RN) black hole embedded in HST. The embedding is different and such configuration can not be obtained as a particular case of our solutions. Therefore, we can only compare the Schwarzschild limit of our RN black holes and the Schwarzschild limit of \cite{Cano:2019ycn}. In the limit $q=0$ and setting $G_N^{(4)} = 1$ we can write the entropy as
\begin{equation}
	S_W = 4 \pi M^2 + \alpha' \, \frac{\pi}{2} \,,
\end{equation}
and we match perfectly equation (6.10) of \cite{Cano:2019ycn} for $p=0$.

\bibliographystyle{JHEP2015}
\bibliography{biblio}


\end{document}